\documentclass[12pt,fleqn]{article}

\usepackage[utf8]{inputenc}
\usepackage[T1]{fontenc}
\usepackage[english]{babel}
\usepackage{lmodern}

\usepackage[DIV12]{typearea}
\usepackage{amsmath,amsfonts,amssymb,amscd}
\usepackage{graphicx}
\usepackage{cite}
\usepackage{mathrsfs}
\usepackage{bbm}
\usepackage{units}
\usepackage[small,loose]{subfigure}  
\usepackage{xcolor}
\usepackage{colortbl} 
\usepackage{pifont}
\usepackage{stmaryrd}
\usepackage{cancel}
\usepackage{multirow} 
\usepackage{rotating} 
\usepackage{diagbox} 
\usepackage{arydshln} 
\usepackage{hyperref}
\usepackage{braket}
\usepackage{enumitem}
\usepackage{xspace}

\newcommand{\Eqref}[1]{Eq.~\eqref{#1}}
\newcommand{\Figref}[1]{Fig.~\ref{#1}}
\newcommand{\Tabref}[1]{table~\ref{#1}}
\newcommand{\Secref}[1]{section~\ref{#1}}
\newcommand{\Appref}[1]{appendix~\ref{#1}}

\newcommand{\eVdist}{\kern-0.06em}

\newcommand{\I}{\mathrm{i}}
\newcommand{\e}{\mathrm{e}}

\newcommand{\U}[1]{\ensuremath{\mathrm{U}(#1)}}
\newcommand{\Z}[1]{\ensuremath{\mathbbm{Z}_{#1}}} 
\newcommand{\A}[1]{\ensuremath{\mathrm{A}_{#1}}}

\newcommand*{\rep}[2][]{\ensuremath{{\boldsymbol{#2}#1}}} 

\renewcommand{\bar}[1]{\overline{#1}}
\newcommand{\Id}{\ensuremath{\mathbbm{1}}}

\newcommand{\Sfour}{\ensuremath{S_4}}
\newcommand*{\Dff}[0]{\ensuremath{\Delta(54)}\xspace}
\newcommand*{\Dts}[0]{\ensuremath{\Delta(27)}\xspace}
\DeclareMathOperator{\Out}{Out}
\newcommand*{\Equi}[0]{\ensuremath{E}\xspace}

\numberwithin{equation}{section}
\numberwithin{table}{section}
\allowdisplaybreaks[1]

\definecolor{darkgreen}{HTML}{009900}

\def\mytitle{Symmetries of symmetries \\[0.2cm] and geometrical CP violation}
\title{\mytitle}

\begin{document}

\begin{titlepage}

\setcounter{page}{0}

\begin{flushright}
 TUM-HEP 980/15\\
 FLAVOUR(267104)-ERC-89\\
\end{flushright}

\vspace*{1.0cm}

\renewcommand*{\thefootnote}{\fnsymbol{footnote}}
\begin{center}
{\Large\textbf{\mytitle}}
\renewcommand*{\thefootnote}{\arabic{footnote}}

\vspace{1cm}

\textbf{Maximilian Fallbacher$^a$\,\footnote[1]{Email: \texttt{m.fallbacher@tum.de}}{}
and 
Andreas Trautner$^{a,b}$\,\footnote[2]{Email: \texttt{andreas.trautner@tum.de}}{} 
}
\\[3mm]
\emph{\small
{}$^a$\,Physik Department T30, Technische Universit\"at M\"unchen,\\
~~James--Franck--Stra\ss e, 85748 Garching, Germany
}
\\[3mm]
\emph{\small
{}$^b$\,Excellence Cluster Universe, \\ Boltzmannstra\ss e 2, 85748 Garching, Germany
}
\end{center}

\vspace{1cm}

\begin{abstract}
We investigate transformations which are not symmetries of a theory but nevertheless leave invariant the set of all symmetry elements and representations.
Generalizing from the example of a three Higgs doublet model with \Dts symmetry, we show that the possibility of such transformations signals physical degeneracies in the parameter space of a theory.
We show that stationary points only appear in multiplets which are representations of the group of these \mbox{so--called} equivalence transformations.
As a consequence, the stationary points are amongst the solutions of a set of homogeneous linear equations.
This is relevant to the minimization of potentials in general and sheds new light on the origin of calculable phases and geometrical CP violation.
\end{abstract}

\end{titlepage}

\section{Introduction}

Spontaneous symmetry breaking (SSB) is known to describe a wide range of physical phenomena in Nature ranging from solid state physics to the origin of elementary particle masses.
Independently of the details of a model, the philosophy of SSB is always the same: one demands the conservation of a symmetry at a high scale 
but has some scalar degree of freedom which, at a lower scale, obtains a vacuum expectation value (VEV) which does not preserve the symmetry.

It is conceivable that SSB also plays a vital role for the experimentally verified but poorly understood violation of the combined symmetry of charge conjugation and parity (CP) \cite{Lee:1973iz}.
A particularly outstanding model of spontaneous CP violation is the so--called geometrical CP violation suggested by Branco, Gérard, and Grimus \cite{Branco:1983tn}.
In their three Higgs doublet model (3HDM) with a discrete \Dts symmetry, the CP violating relative phases of VEVs are independent of the exact values of couplings and follow, thus, solely from the underlying symmetry of the model.

Even though several efforts have been undertaken 
to better understand the origin of geometrical CP violation in the original model 
\cite{deMedeirosVarzielas:2011zw,Degee:2012sk,Holthausen:2012dk}, for potentials of higher order \cite{Varzielas:2012nn}, and in multi--Higgs models \cite{Varzielas:2012pd,Ivanov:2013nla}, we think it is fair to say that a complete understanding of the origin of the calculable phases has not yet been achieved.
Nevertheless, there exist models employing geometrical CP violation based on \Dts which include also complete quark \cite{Bhattacharyya:2012pi,Varzielas:2013sla} 
or lepton sectors \cite{Ma:2013xqa, Varzielas:2013eta}. 
Despite the fact that these models typically have difficulties in producing realistic masses and mixing angles \cite{Felipe:2014zka} we think that, 
in principle, geometrical CP violation and the origin of calculable phases is still a feature worth investigating.
 
Charge conjugation and parity are special because they are not internal symmetries in the conventional sense, rather, C and P are outer automorphisms of the symmetries which are present in a model. 
This is true for the Poincar\'{e} group \cite{Buchbinder:2000cq} as well as for continuous \cite{Grimus:1995zi} and discrete \cite{Holthausen:2012dk} internal symmetries. 
Thus, understanding outer automorphisms is essential to understand CP.
For instance, understanding the interrelation of CP and discrete Groups \cite{Holthausen:2012dk, Feruglio:2012cw} has enabled the discovery 
that settings based on certain discrete groups preclude CP symmetries altogether \cite{Chen:2014tpa}.
Despite their relevance to Nature, however, outer automorphisms have generally not received a lot of attention in the literature. 
To the best of our knowledge it has, for example, not been discussed of what relevance outer automorphisms are that are not C or P transformations.

In general, outer automorphisms are transformations that leave invariant the set%
\footnote{For clarity, note that leaving invariant a set does not mean leaving invariant each of its parts. Rather, it is also possible that the individual parts are permuted.} 
of all elements of a given symmetry -- without being themselves part of the symmetry. 
The latter implies that outer automorphisms also interchange representations of the symmetry.
Hence it is clear that for a given model only those outer automorphisms of the symmetry group are relevant which leave invariant the set of all present representations.
Transformations for which this is true will be called equivalence transformations in this work.
The set of equivalence transformations contains as subsets CP transformations and, as the trivial case, also symmetry transformations.
Despite that, there may be other non--trivial, non--CP equivalence transformations.

Investigating equivalence transformations, we will show that they relate different regions of the parameter space of a theory by isomorphisms. 
From this we will conclude that whenever a theory allows for equivalence transformations, there are different regions in the parameter space which give rise to equivalent physical predictions.
Furthermore, we will show that stationary points of potentials always form multiplets under the group of equivalence transformations.
This allows us to derive a set of homogeneous linear equations which constrain the form, i.e.\ the direction and phases, of all stationary points.
 
This work is organized as follows. In \Secref{sec:geometricalCPV} we review the 3HDM with \Dts symmetry of Branco et al.\ \cite{Branco:1983tn} that we use as an example throughout the work.
We discuss parameter space degeneracies in \Secref{sec:redundant} and derive the complete set of all equivalence transformations for our example in \Secref{sec:Delta27ET}. 
Section~\ref{sec:spontaneousCPV} contains a proof that CP truly is spontaneously violated in the \Dts model. 
After understanding the action of equivalence transformations on stationary points in \Secref{sec:ETonVEVs}, we derive necessary conditions on the stationary points in \Secref{sec:calculableVEVs}.
In the appendix we give the complete traditional minimization of the \Dts 3HDM potential, group theoretical details of \Dts and \Dff, as well as some computational details.

\section{The 3HDM potential with \texorpdfstring{$\boldsymbol{\Dts}$}{Delta(27)} symmetry}
\label{sec:geometricalCPV}

Let us discuss the three Higgs doublet model with \Dts symmetry \cite{Branco:1983tn, deMedeirosVarzielas:2011zw}.
In this model one assigns three electroweak Higgs doublets to the three dimensional representation $(H_1,H_2,H_3)=H=\rep 3$ of \Dts. 
Due to the continuous symmetries and the representation content of the model, the actual discrete symmetry group of the Higgs potential is not \Dts but the larger \Dff \cite{deMedeirosVarzielas:2011zw,
Varzielas:2012nn,Ivanov:2012ry,Ivanov:2012fp}.
We will, thus, work with the full discrete symmetry of the potential, $G=\Dff$. 
The conclusions we obtain are, however, completely independent of whether the analysis of the potential is based on \Dts or \Dff. The reason for this as well
as the group theoretical details of \Dff are given in \Appref{app:Delta27}.

Let us again stress that our aim here is not to provide a realistic model but to use the \Dts potential as playground to explore the meaning of outer automorphism 
transformations and the origin of geometrical CP violation.

The complete renormalizable scalar potential which is invariant under the given symmetries can be written as\footnote{This is equivalent to \cite[Eq.\ (14)]{Ivanov:2012ry}, 
where the potential is given in a slightly rearranged form.}
\begin{equation}\label{eq:HiggsPotential}
\begin{split}
 V~=~&V_0+V_I \\
   =~&-m^2\,H_i^\dagger H_i + \lambda_1\left(H_i^\dagger H_i\right)^2 + \lambda_2\left(H_i^\dagger H_i\right)\left(H_j^\dagger H_j\right) 
   + \lambda_3\left(H_i^\dagger H_j\right)\left(H_j^\dagger H_i\right) \\ 
   & +\tilde{\lambda}_4\left[ \left(H_1^\dagger H_2\right)\left(H_1^\dagger H_3\right) + \text{cyclic} \right] + {\rm h.c.}\;.
\end{split}
\end{equation}
The indices $i$ and $j$ run from 1 to 3 with $i\neq j$.
We parametrize the coupling of the phase dependent part of the potential $V_I$ as $\tilde{\lambda}_4=\e^{\I\,\Omega}\,\lambda_4$ with $\lambda_4>0$ and $0\leq\Omega<2\pi$.
All other couplings are real and chosen such that the potential is bounded below. 
In our discussion we will always assume that the vacuum preserves electric charge and, therefore, parametrize the VEVs as 
\begin{equation}\label{eq:VEVparametrization}
 \langle H_i \rangle \, :=\,\begin{pmatrix} 0 \\ v_i\,\e^{\I\,\varphi_i} \end{pmatrix}\;,
\end{equation}
with $v_i>0$ and $0\leq\varphi_i<2\pi$.
Details on the allowed parameter regions, the complete analytical minimization of the potential as well as a proof of validity for the assumption of charge conservation are given in \Appref{app:potential}.

Earlier analyses have shown that this potential gives rise to very specific stationary points, henceforth also called VEVs, with discrete physical phases \cite{Branco:1983tn,deMedeirosVarzielas:2011zw}.
A careful analysis of the potential \eqref{eq:HiggsPotential} (cf.\ \Appref{app:potential}, and also \cite{Varzielas:2012nn,Ivanov:2014doa}) shows that possible VEVs are given by 
\begin{equation}\label{eq:VEVs}
\langle H \rangle ~=~ v_{\rm I} ~=~ v \begin{pmatrix} 1 \\ 1 \\ 1 \end{pmatrix}\;, \quad
v_{\rm II} ~=~ v \begin{pmatrix} \omega \\ 1 \\ 1 \end{pmatrix}\;, \quad
v_{\rm III} ~=~ v \begin{pmatrix} \omega^2 \\ 1 \\ 1 \end{pmatrix}\;, \quad
v_{\rm IV} ~=~ v \begin{pmatrix} \sqrt{3} \\ 0 \\ 0 \end{pmatrix}\;,
\end{equation}
where here and in the following we use $\omega:=\e^{2\pi\,\I/3}$. 
Each of these four different VEVs actually corresponds to a set of physically equivalent VEVs (a group orbit) which can be obtained by acting on the given vectors with all 
available symmetry transformations. In particular, the overall phase of each VEV is undefined since it can always be shifted by a global hypercharge rotation.

The presence of CP violating physical phases which are independent of couplings and calculable as a consequence of the assumed symmetries is 
called geometrical CP violation \cite{Branco:1983tn}.
In the chosen basis one is easily convinced that CP may be violated spontaneously by the relative geometrical phases of the Higgs VEVs of types $\rm{II}$ and $\rm{III}$ in \eqref{eq:VEVs}. 
Even though less apparent, in \Secref{sec:spontaneousCPV} we will show that also VEVs of the types $\rm{I}$ and $\rm{IV}$ can give rise to spontaneous geometrical CP violation.
While not being overly important to claim that there is geometrical CPV in the first place \cite{Branco:1983tn}, 
realizing that there are the four possible types of VEVs stated in \Eqref{eq:VEVs} is absolutely necessary for the understanding of its origin,
as will become clear at the end of our discussion.

\begin{figure}[t]
\begin{minipage}[]{0.60\linewidth} 
\flushleft
\includegraphics*[width=0.9\textwidth]{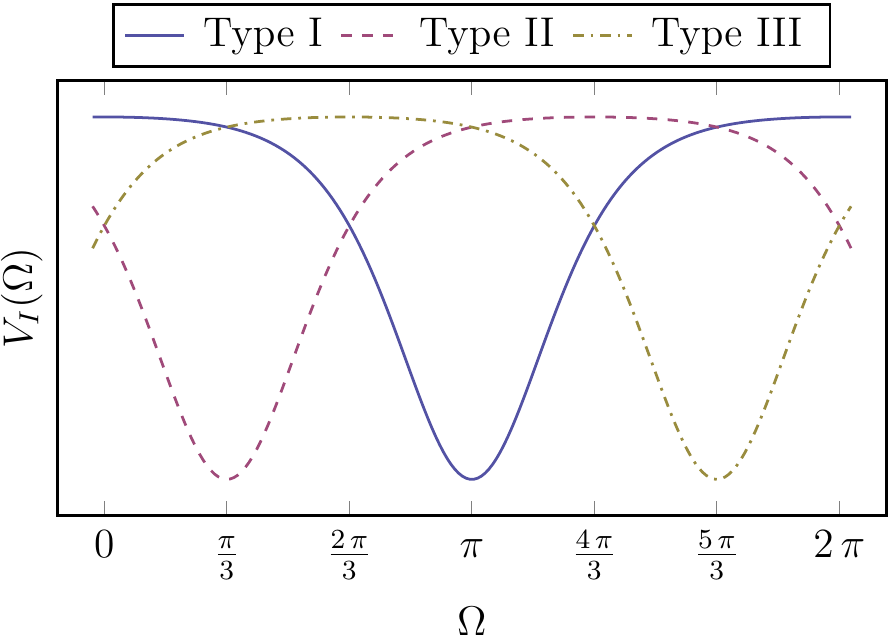} 
\end{minipage}%
\begin{minipage}[]{0.4\linewidth} 
\vspace{-27pt}
\flushright{
\includegraphics*[width=0.9\textwidth]{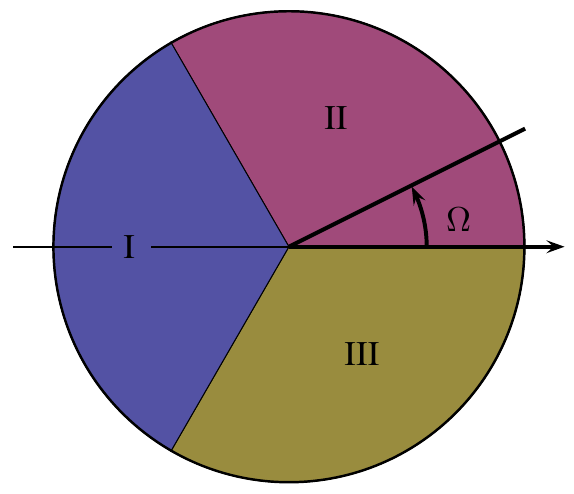} }
\end{minipage}
\caption{Value of the phase dependent potential $V_I$ at the stationary points of type $\rm{I}$, $\rm{II}$, and $\rm{III}$ in dependence of $\Omega$ (left). 
For this illustration, we have chosen couplings $\lambda_{1-4}$ in a region which excludes global minima of type IV.
Note that for the values $\Omega\in\{0,~2\pi/3,~4\pi/3\}$ there are degenerate global minima of two types, whereas for all other values of $\Omega$ the type of the global minimum is unique.
Which stationary point is the global minimum depends discretely on $\Omega$ (right).}
\label{fig:potential}
\end{figure}

Note that the length of the VEV depends on the couplings, $v=v(m^2,\lambda_1,\lambda_2,\lambda_3,\lambda_4,\Omega)$, and has to be determined for each type of VEV individually (cf.\ \Appref{app:potential}).
Thus, which of the stationary points listed in \eqref{eq:VEVs} is the global minimum of the Higgs potential depends on the specific values of the couplings. 
For example, keeping the $\lambda_\ell$ with $\ell=1,..,4$ fixed to values such that a VEV of type $\rm{IV}$ is excluded as global minimum\footnote{This is 
the case which implicitly has been assumed in the analyses of earlier works \cite{Branco:1983tn, deMedeirosVarzielas:2011zw}.} (cf.\ \eqref{eq:minimum_condition}), 
the only parameter which determines the direction of the global minimum is $\Omega$. Whether the global minimum is of type $\rm{I}$, $\rm{II}$, 
or $\rm{III}$ then only depends \textit{discretely} on $\Omega$. This situation is depicted in \Figref{fig:potential} .
We observe that the actual dependence on $\Omega$ is more subtle than only a simple dependence on the sign of $\tilde{\lambda}_4$. 
This clarifies that geometrical CPV also occurs in case $\Omega\notin\{0,\pi\}$, i.e.\ for manifestly complex couplings. Also, this substantiates the statement that the direction of the VEV, 
and especially the relative phases, are stable under renormalization group (RG) running \cite{Branco:1983tn}.
The type of VEV could only change if the parameter evolution were such that (i) $\Omega$ crossed any of the critical values $0$, $2\pi/3$, or $4\pi/3$ or (ii)
the $\lambda_\ell$'s were such that a VEV of type $\rm{IV}$ becomes the global minimum.
Since for both of these conditions the parameter evolution would have to cross coupling values which give rise to an enhanced symmetry, none of them can be fulfilled and the directions of the VEVs are absolutely stable.

\section{Identifying redundant parameter regions}
\label{sec:redundant}

In the following, we want to go beyond the straightforward but rather tedious manual minimization of the potential and discuss the discrete dependence on couplings from a somewhat different point of view. 

For this, note that the parameter space of the potential is partitioned into several regions which differ by the type of VEV which constitutes the global minimum. 
The apparently different types of VEVs, however, all conserve subgroups which are isomorphic. This is true not only for the continuous and discrete internal symmetries of the potential
but also with respect to potential (generalized) CP symmetries as we will see. 
We will show that this is not a coincidence but actually a consequence of the fact that all different parameter regions are redundant in their phenomenology and, thus, 
can be considered physically equivalent -- in a sense that we will specify.

Before we generalize our discussion, let us provide an explicit example to illustrate our point.

The potential \eqref{eq:HiggsPotential} is, in general, a function of field variables and couplings, $V=V(H,\lambda)$, 
where $H$ denotes the fields and $\lambda\equiv\left\{m, \lambda_1, \lambda_2, \lambda_3, \lambda_4, \Omega \right\}$ the couplings, respectively. 
The functional form of $V$ is fixed by the form of all symmetry elements $\rho(\mathsf{g})$ with $\mathsf{g}\in G$, 
since it is required to fulfill $V(\rho(\mathsf{g})H,\lambda)=V(H,\lambda)$.
In our particular case, this function $V(H,\lambda)$ has in addition the striking property that certain transformations performed on \emph{either} the field variables $H$ \emph{or} on the parameters 
$\lambda$ give rise to the same result.\footnote{We explicitly talk about the limited set of scalar parameters $\lambda$ here, so this is no general basis transformation.}
Consider for example a transformation on the Higgs triplet $H\mapsto U_1H$ with the unitary matrix 
\begin{equation}\label{eq:example_trafo}
 U_1~=~\begin{pmatrix}
 \omega & 0 & 0 \\
 0 & 1 & 0  \\
 0 & 0 & 1 \\
 \end{pmatrix}\;.
\end{equation}
Performing this transformation on the potential does not change the functional dependence of $V$ on $H$, i.e.\ it does not change the form of any of the present operators.
In particular, $H\mapsto U_1H$ leaves invariant the mass term, the quartic operators with coupling $\lambda_1$, $\lambda_2$, and $\lambda_3$, as well as 
the value of $\lambda_4$. The only effect of the transformation is a shift of the parameter 
$\Omega$ to a value $\Omega+2\pi/3$. Therefore, this transformation has the only effect of moving the theory to a different point in parameter space, or formally written
\begin{equation}\label{eq:equivalenceCondition}
 V(H',\lambda)~=~V(H,\lambda')\;,
\end{equation}
where we use $H'=U_1H$ and $\lambda'$ to denote the shifted fields and parameters, respectively.
It is crucial here that the respective functional dependences of $V$ on $H$ and on $H'$ are exactly the same.

The striking consequence of the existence of the transformation $H \mapsto U_1 H$ is that the a priori completely unrelated potentials $V(H,\lambda)$ and $V(H,\lambda')=V(H',\lambda)$ make exactly the same physical predictions;
in the first case with respect to $H$ and in the second case with respect to $H'$.\footnote{This conclusion has also been reached in \cite{Ivanov:2006yq}.}
This particularly includes all possible CP transformations and possible residual symmetries after the spontaneous breaking of the original symmetry. 

In the sense that they make the same physical predictions but for two differently defined sets of fields, we say that the theory $V(H,\lambda)$ is equivalent to the theory $V(H,\lambda')$.
Any transformation $U$, which relates two -- in this sense equivalent -- parameter regions, is termed equivalence transformation.

We see that equivalence transformations can be used to relate different regions of parameter space. 
For the complete discussion of the physical phenomenology of a model it is, thus, sufficient to consider only a confined region of the parameter space 
which is related to the complete parameter space by equivalence transformations.

Even though it is always possible to perform the according field redefinitions we think a comment is in order.
Since it is, in principle, possible to distinguish the different components of a triplet, say $H_i$ and $H_{j\neq i}$, from one another by appropriate measurements with respect to subgroups of \Dff, 
it is also possible to distinguish $H_i$ from a corresponding state $H'_i=(UH)_i$ \cite{Chen:1985zzd}.
Therefore, the equivalence property of $V$ can be used to reduce the size of the parameter space only if one does not insist on a relation
between physical states and field operators to begin with.
For example, if one is to determine the parameter $\Omega$ by a measurement one can \emph{either} start by defining 
the state $H_1$ and then has to allow for all values of $\Omega$ \emph{or} choose to describe the measurement 
within a confined region of the parameter space, say $\Omega\in\left[0,2\pi/3\right)$, but then has to relabel the states according to the measured result. 
With an appropriate labeling of states it will never be necessary to leave the constrained parameter range.

For clarity, let us also comment on the relation of equivalence transformations to (Higgs--)basis changes, which are also sometimes called reparametrization transformations \cite{Botella:1994cs,Ivanov:2006yq,Ivanov:2010ww}.
Since the physical results of a theory do, of course, not depend on the specific way the Lagrangean is expressed, it is always possible to perform a field redefinition, i.e.\ to rewrite 
the Lagrangean in terms of new fields $\widetilde{H} = UH$ with an arbitrary unitary matrix $U$.
The resulting potential
\begin{equation}
  \widetilde{V}(\widetilde{H},\lambda):= V(U^{-1}\widetilde{H},\lambda)
  \;,
\end{equation}
however, is in general a different function of its arguments than $V$. Consequently it is, in general, 
impossible to pass on the difference in the functional dependence of $\widetilde{V}$ in comparison to $V$ to the couplings $\lambda=\left\{m, \lambda_1, \lambda_2, \lambda_3, \lambda_4, \Omega \right\}$.
This is possible if and only if $U$ is an equivalence transformation, in which case we have
\begin{equation}
 \widetilde{V}(\widetilde{H},\lambda)~=~V(\widetilde{H},\widetilde{\lambda})\;.
\end{equation}

As just pointed out, the crucial point for any equivalence transformation is that it does not change the functional form of $V$. Because the form of $V$, as discussed above, is fixed by the form of 
all the symmetry elements, this is equivalent to saying that an equivalence transformation must leave invariant the set of all symmetry elements, i.e.\ in our case the set of all matrices of the triplet representation of \Dff generated by $\{A,B,C\}$.
Since this is achieved only by transformations which are automorphisms of the total symmetry group of the potential $G$, it is clear that equivalence transformations have to be automorphisms of $G$.
Since inner automorphisms are by definition induced by the symmetry elements themselves, they act trivially on the couplings. Therefore, we conclude that all non--trivial equivalence transformations
are outer automorphisms of $G$.
Conversely, which of the outer automorphisms of $G$ are equivalence transformations critically depends on which representations of $G$ are present in a specific model.

In general, for an outer automorphism which acts as $u: \mathsf{g}\mapsto u(\mathsf{g})$ and maps a representation $\rep r$ to a representation $\rep{r'}$, 
the explicit representation matrix $U$ is given by the solution to
\begin{equation}\label{eq:consistencyEquation}
U\rho_{\rep{r'}}(\mathsf{g})U^{-1}=\rho_{\rep{r}}(u(\mathsf{g}))\;,\qquad\forall \mathsf{g}\in G\;,
\end{equation}
where $\rho_{\rep r}(\mathsf{g})$ denotes the matrix representation of $\rep{r}$.\footnote{Note that, therefore, the matrices $U$ are 
always defined only up to a phase which is consistent with the fact that a global rephasing of any state cannot matter physically.}
Hence, it is possible, that certain outer automorphisms leave invariant the set of all representation matrices present in a given theory.

This is easily confirmed for the example given above, where $U_1$ in \eqref{eq:example_trafo} is the explicit representation of the automorphism $(\mathsf{A,B,C})\mapsto(\mathsf{BAB,B,C})$ 
acting on, and leaving invariant, the set of all triplet representation matrices.
More generally, consider the case that the symmetry group of a specific model allows for an outer automorphism which maps $\rep r\mapsto\rep r$ for all representations present without being a symmetry of the theory.
Then such a model will unavoidably face degeneracies in the parameter space due to this equivalence transformation. 

Another example is a (possibly generalized) CP transformation for which all representations of $G$ that are used in a model are mapped to their complex conjugate representations $\rep r\mapsto\rep r^{*}$.\footnote{CP
transformations are representations of the outer automorphism group of a discrete symmetry \cite{Holthausen:2012dk} which map
all present representations to their complex conjugate representation \cite{Chen:2014tpa}.} 
If this transformation is not a symmetry, it is well--known that it will map the theory to a different spot in the parameter space. 
The resulting theory is equivalent to its pre--image in the sense that it describes the same dynamics as before but for the CP conjugate set of fields.
Whether we describe the underlying physics with fields or their respective conjugates, however, is completely arbitrary at the level of our example potential, such that 
we may as well work within a restricted region of the parameter space and decide the latter a posteriori.
As soon as we have defined a measurement to tell apart fields from their CP conjugates, however, we would have to consider the complete parameter space.

Note that the only difference between the first and the second example is that in the first case, $H$ and $UH$ transform in the same representation with respect to any further symmetries, whereas in the second case $H$ and $UH^*$ will, in general, transform in complex conjugate representations of any further symmetries.
Thus, the difference between $H$ and $UH^*$ may be defined from elsewhere, whereas the distinction of $H$ and $UH$ can only be made with respect to subgroups of $G$ itself.

\section{Equivalence transformations and the \texorpdfstring{$\boldsymbol{\Dff}$}{Delta(54)} potential}
\label{sec:Delta27ET}

Let us now perform a detailed analysis of the \Dff Higgs potential considering the complete outer automorphism group.
It is most convenient to investigate the action of the outer automorphism group on the potential in a parametrization which is derived directly 
from the $G$--invariant contractions in the chosen basis. Furthermore, we will neglect the mass term because it is invariant under all outer automorphisms 
and focus only on the quartic couplings of the triplet of Higgs doublets $H$.
The contraction \mbox{$(\rep{\bar{3}}\otimes\rep{3})\otimes(\rep{\bar{3}}\otimes\rep{3})$} gives rise to five independent invariants which are given by
\begin{equation}\label{eq:Invariants}
\begin{split}
  \left[\left(H^\dagger_{\rep{\bar{3}}} \otimes H_{\rep{3}} \right)\right.&\otimes\left.\left(H^\dagger_{\rep{\bar{3}}} \otimes H_{\rep{3}} \right)\right]_{\rep[_0]{1}}~=~ 
 a_0\, \left[\left(H^\dagger\otimes H\right)_{\rep[_0]{1}}\otimes\left(H^\dagger\otimes H\right)_{\rep[_0]{1}}\right] \\
+\frac{a_1}{\sqrt{2}}\, &\left[\left(H^\dagger\otimes H\right)_{\rep[_1]{2}}\otimes\left(H^\dagger\otimes H\right)_{\rep[_1]{2}}\right]_{\rep[_0]{1}}  
+\frac{a_2}{\sqrt{2}}\, \left[\left(H^\dagger\otimes H\right)_{\rep[_3]{2}}\otimes\left(H^\dagger\otimes H\right)_{\rep[_3]{2}}\right]_{\rep[_0]{1}} \\
+\frac{a_3}{\sqrt{2}}\, &\left[\left(H^\dagger\otimes H\right)_{\rep[_4]{2}}\otimes\left(H^\dagger\otimes H\right)_{\rep[_4]{2}}\right]_{\rep[_0]{1}} 
+\frac{a_4}{\sqrt{2}}\, \left[\left(H^\dagger\otimes H\right)_{\rep[_2]{2}}\otimes\left(H^\dagger\otimes H\right)_{\rep[_2]{2}}\right]_{\rep[_0]{1}}\;,
\end{split}
\end{equation}
where the $a_k$ for $k=0,..,4$ denote five coupling parameters which can be chosen real since all of the contractions are real themselves.
This parametrization can be compared with the one given in \eqref{eq:HiggsPotential} and we obtain
\begin{equation}\label{eq:parametrization}
\begin{split}
   3\,\lambda_1 ~&=~ a_0+a_4\;,\quad   3\,\lambda_{2} ~=~ 2a_0-a_4\;,\quad 3\,\lambda_3 ~=~ a_1+a_2+a_3\;,  \\
   3\,\lambda_4 ~&=~ \left|a_1+\omega^2\,a_2+\omega\,a_3\right|\;, \quad  \text{and} \quad \Omega ~=~ \arg\left(a_1+\omega^2\,a_2+\omega\,a_3\right)\;.
\end{split}
\end{equation}
Of course, the parameters $a_k$ are subject to constraints due to the physicality of the potential and the form of the vacuum we want to obtain
completely analogous to the constraints on $\lambda_\ell$ (cf.\ \eqref{eq:physicality} and \eqref{eq:physicalityA}).

The complete outer automorphism group of \Dff is $\Sfour$, the permutation group of four elements. This group can be generated%
\footnote{Elements of the outer automorphism group are by definition not automorphisms but cosets of automorphisms. 
To generate the group we therefore have to choose one representative of a coset (all other elements of the coset can be obtained by composition with an inner automorphism).
The results must not depend on the particular choice.}
by two outer automorphisms of order two and three which fulfill\footnote{Note
that $\text{``id''}$ here is not strictly the identity map but may also be an inner automorphism, i.e.\ an element out of the symmetry group \Dff.
This is indeed the case here and $(t^{-1}\circ s)^4=\operatorname{conj}(\mathsf{C})$ only closes to conjugation with the group element $\mathsf{C}$.\label{fn:innerAut}}
\begin{equation}\label{eq:Delta54Algebra}
 t^3 ~=~ s^2 ~=~ \left(t^{-1}\circ s \right)^4~=~\text{id}\;,
\end{equation}
and act on the doublet and triplet representations of \Dff as
\begin{subequations}\label{eq:OutGensDelta54}
\begin{align}
&t:~(\mathsf{A,B,C})\mapsto(\mathsf{A,ABA,C})\,&
&\hspace{-0.5cm}\curvearrowright\,&
&\begin{pmatrix} \rep[_1]{2} \\ \rep[_2]{2} \\ \rep[_3]{2} \\ \rep[_4]{2} \end{pmatrix} \mapsto 
 \begin{pmatrix} \rep[_1]{2} \\ \rep[_4]{2} \\ \rep[_2]{2} \\ \rep[_3]{2} \end{pmatrix}\;,
&
&\quad\rep[_i]{3}\mapsto U_{t}\,\rep[_i]{3}\;,& \\[0.2cm]
&s:~(\mathsf{A,B,C})\mapsto(\mathsf{AB^2A,B,C})\,&
&\hspace{-0.5cm}\curvearrowright\,&
&\begin{pmatrix} \rep[_1]{2} \\ \rep[_2]{2} \\ \rep[_3]{2} \\ \rep[_4]{2} \end{pmatrix} \mapsto 
 \begin{pmatrix} S_{\rep2}\,\rep[_4]{2} \\ S_{\rep2}\,\rep[_2]{2} \\ \phantom{S_{\rep2}\,}\rep[_3]{2} \\ S_{\rep2}\,\rep[_1]{2} \end{pmatrix}\;,
&
&\quad \rep[_i]{3}\mapsto U_{s}\,\rep[_i]{3}^*\;,&
\end{align}
\end{subequations}
with representation matrices which are given by
\begin{equation}\label{eq:Delta54Aut}
S_{\rep2}~=~\begin{pmatrix} 0 & 1 \\ 1 & 0 \end{pmatrix}\;, \quad 
U_{t}~=~\frac{\I}{\sqrt{3}}\begin{pmatrix} 1 & \omega^2 & \omega^2 \\ \omega^2 & 1 & \omega^2  \\ \omega^2 & \omega^2 & 1 \end{pmatrix}\;, \quad 
U_{s}~=~\begin{pmatrix} \omega^2 & 0 & 0 \\ 0 & 0 & 1  \\ 0 & 1 & 0 \end{pmatrix}\;.
\end{equation}
All other elements of the outer automorphism group can be generated as compositions of $t$ and $s$.

With respect to any of the three--dimensional representations, 
the outer automorphism group of \Dff splits into two kinds of transformations:
\begin{enumerate}[label=(\roman*)]
  \item \label{enum:equiv} Transformations that send $\rep3\to\rep3$. 
  \item \label{enum:CP} Transformations that send $\rep3\to\rep3^*$.
\end{enumerate}
We find that there are $12$ possible transformations of the first kind (three of order two, eight of order three and the identity, i.e.\ all even permutations) and
$12$ possible transformations of the second kind (six of order two and six of order four, i.e.\ all odd permutations). 
The results for the second case are in accordance with the findings of \cite{Nishi:2013jqa}.
Note that any transformation \ref{enum:equiv}, if conserved, would increase the linear symmetry of the theory whereas any transformation \ref{enum:CP} would warrant CP conservation 
(it might, in addition, increase the linear symmetry, too).
If not conserved, both transformations map the theory to different points in parameter space and, hence, are equivalence transformations
on an equal footing. Thus, in our example model the group of equivalence transformations is the complete available outer automorphism group.

From the generators it is straightforward to obtain the corresponding transformations of the couplings. Let us illustrate this with a few examples.

We start with transformations of type~\ref{enum:equiv}.
Take, for example, the transformation $t$.
Acting with it on the triplet in \eqref{eq:Invariants} is equivalent to the parameter mapping 
\begin{equation}\label{eq:P1Action}
 (a_1,a_2,a_3,a_4)\mapsto(a_1,a_3,a_4,a_2)\;.
\end{equation}
This result can be cross--checked also in the conventional form of the Lagrangean \eqref{eq:HiggsPotential}.
The parametrization in form of the derived invariants \eqref{eq:Invariants} is clearly advantageous since the
transformation of the parameters can easily be obtained from the transformation of the doublets in \eqref{eq:OutGensDelta54}. 
The theory with parameters $(a_1,a_2,a_3,a_4)$, hence, is equivalent in the above sense to a theory with $(a_1,a_3,a_4,a_2)$.

The transformation \eqref{eq:example_trafo}, which we have used as an example before, is given by $s\circ t^{-1}\circ s\circ t$ (modulo an inner automorphism with the element $\mathsf{C}$) and is equivalent to the parameter mapping
\begin{equation}
 (a_1,a_2,a_3,a_4)\mapsto(a_2,a_3,a_1,a_4)\;,
\end{equation}
which, if applied to \eqref{eq:parametrization}, confirms that $\Omega\mapsto\Omega+2\pi/3$. Again, this transformation identifies parameter regions which are equivalent in the above sense.

Let us now discuss potential CP transformations \ref{enum:CP}. 
A priori, all transformations which map $\rep3$ to $\rep3^*$ are possible physical CP transformations because each --~if conserved~-- 
ensures that all CP violating (basis--)invariants vanish.
Taking, for example, the explicit action of $s$, it is straightforward to confirm that this transformation is equivalent to the map
\begin{equation}\label{eq:U1Action}
 (a_1,a_2,a_3,a_4)\mapsto(a_3,a_2,a_1,a_4)\;.
\end{equation}
This implies that theory with parameters $(a_1,a_2,a_3,a_4)$ describes, with respect to $H$, precisely the same dynamics as a theory with parameters $(a_3,a_2,a_1,a_4)$ with respect to $U_{s}H^*$.  
Therefore, $s$ is a CP symmetry of the theory if and only if the couplings fulfill the relation $a_1=a_3$.
Requiring this relation in \eqref{eq:parametrization} implies that $\Omega\in\{\pi/3, 4\pi/3 \}$, i.e.\ the phase of the complex coupling is fixed to very specific values
-- just as one would naively expect from a CP transformation.

Another example one might be interested in is what in our basis could be called the canonical CP transformation.\footnote{Note that none of the order two CP transformations here is distinguished with 
respect to the other order two CP transformations. It is rather a matter of basis choice which CP transformation
one would call canonical and which one would call generalized.}
This transformation acts on the triplets as $\rep{3}\mapsto U\rep{3}^*$ with $U=\Id$ and is given by $s\circ t^{-1}\circ s\circ t\circ s$. This is equivalent to the map
\begin{equation}
 (a_1,a_2,a_3,a_4)\mapsto(a_1,a_3,a_2,a_4)\;,
\end{equation}
implying $\Omega\in\{ 0, \pi \}$ if this CP transformation is to be conserved.
The absence of any phases in the Lagrangean is, of course, what one would naively expect if CP is conserved. 
Nevertheless, as seen above (see also \cite{Holthausen:2012dk,Ivanov:2014doa}) this is not the only way in which CP can be conserved physically. Indeed, CP is conserved whenever two of the four parameters $(a_1,a_2,a_3,a_4)$ are equal.

Taking together all possible equivalence transformations, the Higgs potential with \Dff symmetry and a given set of parameters is equivalent to every 
potential which can be obtained by any permutation of the four parameters $a_\ell$.
This equivalence can be made explicit by a field redefinition for all even permutations and a complex field redefinition, 
i.e.\ a CP transformation, for all odd permutations of the $a_\ell$, respectively.
It is noteworthy that the action of these equivalence transformations in the conventional parametrization \eqref{eq:HiggsPotential}
may not always be as simple as just a shift in the phase $\Omega$, as in our example, but can also affect the other parameters.

As a result of this discussion the potential can be analyzed within a restricted region of the parameter space without missing any of its phenomenological features.
In case CP is broken explicitly, a possible choice for a non--degenerate parameter space is \mbox{$a_1<a_2<a_3<a_4$}. In case an order two CP symmetry is required to be conserved initially,  
a possible choice for the restricted parameter region is \mbox{$a_1<a_2<a_3=a_4$} or \mbox{$a_3=a_4<a_1<a_2$}, where in the first case CP is conserved before and after the spontaneous breakdown of $G$, 
while in the second case CP is spontaneously violated. All parameter regions that can be obtained from the three given ones by any permutation of the four couplings are equivalent in the above sense and therefore not explicitly stated.
In contrast, requiring that more than two parameters are equal leads to an enhancement of the linear symmetry of the model.%
\footnote{\label{fot:SymEnhancement}In case we have two pairs of equal parameters or three (or more) equal parameters the discrete symmetry of the potential is enhanced from \Dff to 
$\widetilde{G}:=\left(\left(\Z3\times\Z3\right)\rtimes\Z3\right)\rtimes\Z4\cong\text{SmallGroup(108,15)}$ or to a continuous group, respectively, 
in agreement with the maximal realizable symmetry $\Sigma(36)\cong \widetilde{G}/\Z3$ found in \cite{Ivanov:2012ry,Ivanov:2012fp}.}

\section{Spontaneous (geometrical) CP violation}
\label{sec:spontaneousCPV}

Having at hand the complete automorphism group including all possible CP transformations, let us 
comment on the phenomenon of spontaneous geometrical CP violation first described in \cite{Branco:1983tn}.

Since there has been no discussion of any process which makes the violation of CP tangible,
the question is whether CP really can be violated physically if $H$ assumes any of the VEVs given in \Eqref{eq:VEVs}.
In order to confirm that this is indeed the case, let us assume that a CP transformation acting on the Higgs fields as $H\to UH^*$ with an arbitrary $U$ is a symmetry of the Lagrangean.
Although \Dts does not allow for CP transformations in a generic setting \cite{Chen:2014tpa}, imposing such a transformation here is possible because the model contains only triplet representations.
In order for the chosen CP transformation to be spontaneously broken by the Higgs VEV, the condition
\begin{equation}\label{eq:VEVcondition}
\langle H \rangle ~=~ U\,\langle H \rangle^*\;
\end{equation}
must be violated. 
However, in order to claim that CP is really violated physically by $\langle H \rangle$, 
we have to make sure that there is \emph{no other} CP transformation which is fulfills \eqref{eq:VEVcondition} for the corresponding $U$, while at the same time being a symmetry of the Lagrangean.

Without loss of generality\footnote{As the CP transformations are mapped onto each other under the action of the equivalence transformations, all order two CP transformations are equivalent at the level of the potential.} 
we may focus on the CP transformation induced by the automorphism $s$, which acts on the Higgs triplet with the matrix $U_{s}$ given in \Eqref{eq:Delta54Aut}.
This transformation is a symmetry of the Lagrangean if and only if $\Omega\in\{\pi/3, 4\pi/3 \}$ (corresponding to $a_1\equiv a_3$) as discussed after \Eqref{eq:U1Action}. 
Depending on the values of the other parameters, the VEV can be of any type ${\rm I-IV}$.
It is straightforward to check that \eqref{eq:VEVcondition} with $U=U_{s}$ is fulfilled for VEVs of the types $\rm{II}$ and $\rm{IV}$ and violated for VEVs of the types $\rm{I}$ and $\rm{III}$, respectively.
Thus, in order for $U_s$ to be spontaneously broken, we require that the global minimum is either of type $\rm{I}$ or of type $\rm{III}$ which can only be the case if $\Omega$ is $4\pi/3$.
In order to claim that this also implies the spontaneous violation of CP, however, we have to ascertain that there is no other CP symmetry of the Lagrangean which solves \eqref{eq:VEVcondition}.
In the case at hand, it is straightforward to check that all other possible CP transformations are broken explicitly if we do not allow for any further parameter relations.
In turn, if we allowed for any additional parameter relation, the linear (i.e.\ non--CP) symmetry of the potential would unavoidably be enhanced as well (cf.\ the discussion in footnote \ref{fot:SymEnhancement}).
Therefore, not allowing for an enhancement of the discrete symmetry \Dff, we have shown that there is physical CP violation with calculable phases from the spontaneous breaking of the CP symmetry $s$. 
As the source of CP violation we identify quadratic and cubic couplings of the Higgs potential after the SSB.

In particular for VEVs of the types $\rm{I}$ and $\rm{IV}$ the fact that CP can be violated geometrically may appear surprising because none of the Higgs VEVs assumes a complex phase. 
That our statement is nevertheless correct can be understood by noting that the corresponding CP transformation matrices, in the above example $U_s$, 
carry discrete phases which also enforce discrete non--trivial values of $\Omega$, i.e.\ give rise to calculable discrete phases in the Lagrangean.
Alternatively, one can also use a basis change to bring $U_s$ to the canonical form, thereby shifting the geometrical phases to the VEVs. 
In general, the origin of complex phases in matrices which represent outer automorphisms, and in particular CP transformations, 
can be tracked back to the necessarily complex Clebsch--Gordan coefficients of the symmetry group $\Dff$ \cite{Chen:2014tpa}.

Let us note that from the relation $a_1\equiv a_3$ it follows immediately that vacua $\rm{I}$ and $\rm{III}$ are energy degenerate.
This is also clear because the vacua are part of the same group orbit with respect to the broken CP symmetry generated by $U_s$.
Since the two vacua are in principle distinguishable, there are domain walls present after the spontaneous breaking,
where the different domains then have different properties also with respect to CP.

Let us also comment on the situation that we introduce other sectors, such as Yukawa couplings to fermions, to a model.
If the new sector does not obey the full symmetry group of the Higgs potential but only a smaller group, the discussion of equivalence and CP transformations obviously has to be based on the outer automorphisms of that group.
But even if the symmetry is not reduced, the presence of fields in representations other than the Higgses' typically reduces the number of available equivalence and CP transformations.
This is because some of the equivalence transformations are rendered impossible by the fact that they imply mappings of representations onto other representations which are not present in the model, i.e.\ the corresponding transformations are broken explicitly and maximally.
Therefore, it may happen that VEVs, which are equivalent at the level of the Higgs potential, have different physical implications on masses, mixings, and CP violation in the additional sectors.
This is the case in models which employ the \Dts potential for the explanation of fermion mixing patterns or masses \cite{Bhattacharyya:2012pi,Varzielas:2013sla,Ma:2013xqa, Varzielas:2013eta}.%
\footnote{Contrary to a statement made in \cite{Bhattacharyya:2012pi} it is completely arbitrary which one--dimensional representations of \Dts are chosen for the matter content of their model. 
This is because we can always find a valid CP transformation among the outer automorphisms of \Dts as long as the model contains only one kind of non--trivial singlet representation and its complex conjugate.}
Even though in this case the complete set of all equivalence transformations may not be useful to identify equivalent parameter regions of the full theory, 
it is still a powerful tool to analyze the Higgs potential, i.e.\ to obtain all possible VEVs, as we will show in the following.

\section{Action of equivalence transformations on VEVs}
\label{sec:ETonVEVs}

Let us discuss the action of equivalence transformations on the VEVs of a potential.
For this, assume that the potential $V(H,\alpha$) has a VEV $\phi(\alpha):=\langle H \rangle$ which is, in general, a continuous function of the couplings $\alpha\equiv(m,a_0,a_1,a_2,a_3,a_4)$.
It is well--known that the action of symmetry transformations on a given VEV gives rise to physically equivalent VEVs, which are distinct from the original one if the corresponding 
symmetry transformation is spontaneously broken.
All VEVs which are related by the action of a symmetry transformation are said to lie on a so--called group orbit.

In close analogy to symmetry transformations, the characteristics of equivalence transformations imply that VEVs of the potential are related by certain transformations.
More specifically, it is possible to obtain new VEVs from known ones simply by taking
\begin{equation}\label{eq:additionalVEV}
  \phi'\left(\alpha\right)~=~
  \left\{\begin{array}{ll}
    U\,\phi\left( \alpha' \right)\;,& \text{~or}\\[0.2cm]
    U\left(\phi\left( \alpha' \right)\right)^*\;, & \text{~}
  \end{array}\right.
\end{equation}
where $U$ is the representation matrix of an equivalence transformation, $\alpha'$ the correspondingly transformed couplings, and $\phi'(\alpha)$ denotes a new VEV of the original potential $V(H,\alpha$).
The first line holds if $U$ represents an equivalence transformation which maps the representation of $H$ to itself, whereas the second line refers to transformations which map the representation of $H$ to its complex conjugate.
A proof of \eqref{eq:additionalVEV} is given in \Appref{app:VEVtrafo}.

In case $U$ is a non--trivial equivalence transformation (i.e.\ not a symmetry transformation), the VEVs 
$\phi'(\alpha)$ and $\phi(\alpha)$, which are related by \eqref{eq:additionalVEV}, are, in general, not part of the same symmetry group orbit.
Therefore, orbits of VEVs under the action of non--trivial equivalence transformations are `perpendicular' to the group orbits in the sense that $\phi'(\alpha)$ 
cannot be obtained from $\phi(\alpha)$ by a symmetry transformation.
Hence, \eqref{eq:additionalVEV} can be used to obtain new minima from known ones beyond the group orbit.
For example could one have simply guessed the first VEV in \Eqref{eq:VEVs} and obtained all other VEVs of \Eqref{eq:VEVs} by the application of equivalence transformations, thereby avoiding tedious manual computations.
Also, this shows that stationary points are transformed among themselves via equivalence transformations, i.e.\ they always appear in complete multiplets of the available group of outer automorphisms.
The consequences of this insight will be elucidated in the following.

\section{A necessary condition on the VEVs}
\label{sec:calculableVEVs}

In the previous section we have established that VEVs do not only form orbits under the symmetry group $G$ but also orbits under the full group of equivalence transformations \Equi, 
which also contains certain outer automorphisms of the symmetry group.

Let us in the following put aside the existence of the $\U{1}_\text{Y}$ symmetry, i.e.\ the fact that our VEVs can be re--phased continuously,
and focus on the orbits under the discrete equivalence transformations of the potential.
The maximal number of distinct VEVs that can be obtained from a given VEV by equivalence transformations, i.e.\ the maximal orbit length, is then given by the number of possible discrete transformations $|\Equi|=|G|\times|\Out(G)|$.

However, due to the fact that the VEVs here are, by definition, solutions to a well--behaved system of coupled polynomial equations%
\footnote{`Well--behaved' here is used in the mathematical sense meaning that the system of equations consists of as many equations as variables and has only a finite number of solutions.},
their number is strictly bounded above (cf.\ e.g.\ \cite{Schmid1995}).
If the maximal orbit length $|\Equi|$ exceeds the number of possible solutions, all VEVs have to be fixed points of at least one equivalence transformation and possibly of several more.
This in turn can be used to set up a necessary condition on the VEVs which restricts their possible directions and phases.

We will now use our example model to illustrate this method.
However, it can, of course, be adapted to any other potential and we will comment on this below.
In our example model, the group of equivalence transformations \Equi, which describes the complete orbit of stationary points, can be presented by the five generators $\{\mathsf{A,B,C,S,T}\}$ with the relations
\begin{align}\label{eq:presentationP}\nonumber
 &\mathsf{T}^3~=~\mathsf{S}^2~=~\left(\mathsf{T}^{-1}\mathsf{S}\right)^4~=~\mathsf{e}\;,& \\\nonumber
 &\mathsf{T\,A\,}\mathsf{T}^{-1}~=~\mathsf{A}\;, & \mathsf{S\,A\,S}^{-1}&~=~\mathsf{AB^2A}\;, & \\ 
 &\mathsf{T\,B\,T}^{-1}~=~\mathsf{ABA}\;, & \mathsf{S\,B\,S}^{-1}&~=~\mathsf{B}\;, & \\\nonumber
 &\mathsf{T\,C\,T}^{-1}~=~\mathsf{C}\;, & \mathsf{S\,C\,S}^{-1}&~=\mathsf{C}\;, 
\end{align}
in addition to the relations for \Dff given in \eqref{eq:presentationG}.
In the SmallGroup catalogue of GAP \cite{GAP4}, \Equi is given by $\mathrm{SG}(1296,2891)$ and has order $|\Equi|=|G|\times |\Out(G)|=1296$ as expected.

The number of VEVs, however, is strictly bounded above by \mbox{$3^6=729$} \cite{Schmid1995}.
This number is obtained from the fact that, with respect to the discrete symmetries, $H$ consists of a triplet of complex scalars and the potential is renormalizable, 
thus yielding VEVs which have to be solutions to a system of six coupled polynomial equations of degree three.

Therefore, any \Equi--orbit of VEVs under the \mbox{(left--)action} of \Equi must be of size smaller than $729$ and, hence, smaller than $|\Equi|$. 
This implies that for \emph{any} VEV $\phi$ there must be a non--trivial subgroup of \Equi, denoted by $\Equi_\phi$, which leaves $\phi$ invariant.
This is a direct consequence of the orbit--stabilizer theorem (e.g.\ \cite[p.\ 80]{CeccheriniSilberstein2008}).

In case there are several distinct orbits of VEVs under the action of \Equi they are disjoint. 
Therefore, we can consider each orbit separately.
Let us denote by $\Phi$ an orbit of VEVs corresponding to $\phi$, i.e.\ $\Phi:=\{p\phi \,|\, p \in \Equi\}\equiv\Equi \phi$.
Using the fact that, by construction, $G$ is a normal subgroup of \Equi, one can show (e.g.\ \cite[p.\ 12]{Bogopolskij2008}) that $\Phi$ has the structure

\begin{equation}
 \Phi^\mathsf{T}~=~\begin{pmatrix} \,\boxed{\leftarrow ~ G\,\phi_1 ~ \rightarrow} \,, \,\boxed{\leftarrow ~ G\,\phi_2 ~ \rightarrow}\,, \qquad \cdots \qquad ,  \,\boxed{\leftarrow ~ G\,\phi_n ~ \rightarrow} ~~\end{pmatrix}\;,
\end{equation}
where the boxes denote equally--sized blocks which contain $G$--orbits of VEVs $G\phi_i\equiv \{g\phi_i \,|\, g \in G\}$.
The individual blocks have size \mbox{$r:=|G|/|G\cap \Equi_\phi|$} and the number of blocks is given by $n:=|\Equi||G\cap \Equi_\phi|/\left(|G||\Equi_\phi|\right)$.
The orbit--stabilizer theorem guarantees that $|\Phi|=|\Equi|/|\Equi_\phi|=r\cdot n\,$.

Under the action of elements in $G$, the VEVs are permuted transitively only within the individual blocks, 
whereas under the action of elements in \Equi which are not in $G$ the blocks themselves are permuted transitively.
This is what we mean by calling the $G$--orbits `perpendicular' to the outer automorphism orbits in \Secref{sec:ETonVEVs}.

Let us now investigate in detail how $\Phi$ transforms under \Equi.
On the one hand, the explicit action of \Equi on the triplet VEVs $\phi$ is given by the representation matrices obtained from \Eqref{eq:consistencyEquation}.
On the other hand, it is clear that this action must transform each $\phi\in\Phi$ to another VEV of the same orbit and, thus, must be given by a permutation of the components of $\Phi$.
Because \Equi acts transitively on $\Phi$, this permutation can be shown to be equivalent to the permutation of the elements of the coset space $\Equi/\Equi_\phi$ under 
the action of \Equi by left--multiplication (e.g.\ \cite[p.\ 80]{CeccheriniSilberstein2008}).%
\footnote{The same mathematical equivalence is used in the construction of effective Lagrangeans in presence of spontaneously broken continuous symmetries, where, however, only the action of the symmetry group is considered in order to parametrize the vacua \cite{Coleman:1969sm,Callan:1969sn}.}
The possible sets of VEVs $\Phi$ are, therefore, constrained to the possible permutation representations of $\Equi/\Equi_\phi$ (under left--action of \Equi) for 
all possible subgroups $\Equi_\phi$.
The fact that the explicit action of \Equi on $\Phi$ must correspond to one of these possible permutations imposes a necessary condition on all VEVs with non--trivial stabilizer.
Due to the upper bound on the number of VEVs presented above, all VEVs have a non--trivial stabilizer in our case.

Let us explicitly derive these necessary conditions. 
Instead of working with the five generators of \eqref{eq:presentationP} it is more convenient to work with a minimal set of generators of \Equi 
which can be obtained with GAP and is given by $\{\mathsf{P},\,\mathsf{Q}\}$ with
\begin{equation}
 \mathsf{P}~:=~\mathsf{T}\;, \quad\text{and}\quad 
 \mathsf{Q}~:=~(\mathsf{T\,S})^2\,(\mathsf{T}^{-1}\,\mathsf{S})^2\, \mathsf{C}
 \,(\mathsf{T}^{-1}\,\mathsf{S})^2\, \mathsf{C}\, \mathsf{A}\, (\mathsf{T}^{-1}\,\mathsf{B}^{-1}\,\mathsf{T\,B\,A})^4\;,
\end{equation}
where the action on the triplet representation is given by
\begin{equation}
  P~=~\frac{\I}{\sqrt{3}}\,\begin{pmatrix} 1 & \omega^2 & \omega^2\\ \omega^2 & 1 & \omega^2\\ \omega^2 & \omega^2 & 1\end{pmatrix}\, \qquad \text{and} \qquad Q~=~ \frac{\I}{\sqrt{3}}\,\begin{pmatrix} \omega^2 & \omega & \omega^2\\ \omega & \omega & 1\\ 1 & \omega & \omega\end{pmatrix}\,.
\end{equation}
For convenience, let us work with the representation $\rep{6}=\rep{3}\oplus\rep{\bar{3}}$. This is advantageous because we can simply multiply together $\rep{6}$--plet matrices without having to pay special attention to 
transformations which involve complex conjugation of the triplet. Elements of the $\rep{6}$--plet which involve the complex conjugation of the triplet (e.g.\ $\mathsf{S,Q}$) 
are simply represented by matrices which have non--zero blocks only on the anti--diagonal, whereas all other elements
(e.g. $\mathsf{A,B,C,T\equiv P}$) have a block--diagonal structure (cf.\ e.g.\ \cite{Holthausen:2012dk}).
The representation matrices of the minimal generating set for the \rep{6}--plet representation are
\begin{equation}
  P_{\rep{6}}  ~=~ \begin{pmatrix} P & 0\\ 0 & P\end{pmatrix} \qquad \text{and} \qquad Q_{\rep{6}}~=~\begin{pmatrix} 0 & Q\\ Q & 0\end{pmatrix}\,. 
\end{equation}

Next, we want to obtain the permutation representation of the set of VEVs $\Phi$, for which it is required to assume a certain stabilizer subgroup $\Equi_\phi$.
The minimal generating set of the corresponding permutation matrices, which we denote by $\Pi_{\mathsf{P}}$ and $\Pi_{\mathsf{Q}}$, can for example be obtained via GAP \cite{GAP4}, see \Appref{App:VEVCalculation}.

For consistency now, the direct action of an element of \Equi on \emph{each} $\phi\in \Phi$ must have the same effect as the corresponding permutation acting on \emph{the whole set} $\Phi$.
The representation matrices for these two actions on $\Phi$ are given by
\begin{equation}
 P_\Phi~:=~\bigoplus_{i=1}^{r\cdot n}\, P_{\rep{6}}\;, \qquad \text{and} \qquad Q_\Phi~:=~\bigoplus_{i=1}^{r\cdot n}\, Q_{\rep{6}}\;,
\end{equation}
for the direct transformation of the VEVs and by
\begin{equation}\label{eq:AbsSandT}
 \Pi_{\mathsf{P}}^\Phi~:=~\Pi_{\mathsf{P}}\otimes\Id_6\;, \qquad \text{and} \qquad \Pi_{\mathsf{Q}}^\Phi~:=~\Pi_{\mathsf{Q}}\otimes\Id_6\;,
\end{equation}
for the permutation, respectively. Here $\oplus$ denotes the matrix direct sum and $\otimes$ the Kronecker product of matrices.
Consistency now requires that
\begin{equation}\label{eq:consistency}
 \left(P_\Phi-\Pi_{\mathsf{P}}^\Phi\right)\,\Phi~=~0\;, \qquad \text{and} \qquad  \left(Q_\Phi-\Pi_{\mathsf{Q}}^\Phi\right)\,\Phi~=~0\;.
\end{equation}
These two homogeneous linear equations are fulfilled by the orbit $\Phi$ of any VEV $\phi$ with stabilizer $\Equi_\phi$.
Turning this around, it is possible to find candidates for VEVs by assuming a certain $\Equi_\phi$ and then checking for possible solutions to \eqref{eq:consistency}.
Depending on the specific subgroup that is assumed, the combined rectangular matrix
\begin{equation}\label{eq:Mconsitency}
M:=\begin{pmatrix}P_\Phi-\Pi_{\mathsf{P}}^\Phi\\ Q_\Phi-\Pi_{\mathsf{Q}}^\Phi\end{pmatrix}\,
\end{equation}
may either have $\operatorname{rank}(M) = 6\,|\Phi|$, implying that there is only the trivial solution for $\Phi$, or
$\operatorname{rank}(M)<6\,|\Phi|$, implying that there is a non--trivial solution for $\Phi$.
In the first case, VEVs which conserve the assumed subgroup $\Equi_\phi$ cannot exist, whereas in the second case,
the solutions to \eqref{eq:consistency} are candidates for orbits of non--trivial VEVs.

Note that the only information used up to this point is the discrete symmetry group of the potential and the group of available equivalence transformations, i.e.\ information about the representation content of the model.
Therefore, our constraint on the VEVs is independent of the precise form of the potential and simply reveals what (orbits of) VEVs are possible in principle.%
\footnote{In case the orbit of VEVs is allowed to be of the full length $|\Equi|$, one may assume a set $\Phi$ with a trivial stabilizer subgroup.
In this case, the solution to \eqref{eq:consistency} has as many free parameters as $\phi$ has components such that our equations do not impose any constraint on the VEVs.}
In order to check which of the non--trivial solutions to \eqref{eq:consistency} really is a stationary point of the potential and to fix remaining free parameters, 
one still has to plug an element of $\Phi$ into the gradient of the potential.

Performing a scan over all subgroups of \Equi while checking for non--trivial solutions of \eqref{eq:consistency}, we find that, up to conjugation%
\footnote{It is sufficient to limit the scan to conjugacy classes of subgroups due to the fact that stabilizer subgroups of points on the same orbit are conjugate to each other.},
the largest subgroups of \Equi which allow for a non--trivial $\Phi$ are two groups of order 18 ($\mathrm{SG}(18,4)$ and $\mathrm{SG}(18,3)$) and a group of order 48 ($\mathrm{SG}(48,29)$).
Of course, also the subgroups of these groups allow for non--trivial solutions to \eqref{eq:consistency} which, however, are less restrictive on $\Phi$. We will discuss this issue below.
The permutation representations of $\Phi$ corresponding to the largest subgroups are labeled as $\rep{72}_1$, $\rep{72}_2$, and $\rep{27}$, and the corresponding generators 
can be found in \mbox{Appendix \ref{app:permutations}}.

Explicitly solving \eqref{eq:consistency} for the representation $\rep{72}_1$ results in 
\begin{equation}
\Phi_{\rep{72}}^\mathsf{T}~=~ \begin{pmatrix} ~\boxed{ G\,\phi_1 } \,, \,\boxed{ G\,\phi_2 }\,, \,\boxed{ G\,\phi_3 } \,, \,\boxed{ G\,\phi_4 }  ~~\end{pmatrix}\;,
\end{equation}
where $\phi_{1-4}$ are representatives of the different blocks, for example, given by
\begin{equation}\label{eq:Block_representants72}
 \left( \phi_1, \phi_2, \phi_3, \phi_4 \right)~=~\left( \begin{pmatrix}-\omega \\ -\omega \\ -\omega \end{pmatrix}v_1, \begin{pmatrix}-\omega \\ -1 \\ -1 \end{pmatrix}v_2, 
 \begin{pmatrix}\omega \\ \omega^2 \\ \omega^2 \end{pmatrix}v_3, \begin{pmatrix} \I\omega\sqrt{3} \\ 0 \\ 0 \end{pmatrix}v_4\right)\;.
\end{equation}
Modulo a global re--phasing, which is allowed because of the so-far neglected $\U{1}_\text{Y}$, this precisely reproduces the four types of VEVs ${\rm I-IV}$ \eqref{eq:VEVs} found in the conventional way. 
The functions $v_{\ell}$, as well as the type of the VEVs, then can easily be obtained by plugging the $\phi_{\ell}$ into the gradient of the potential. 
We find that, for certain parameter regions, all these VEVs are local minima with
\begin{equation}\label{eq:VEVstrength}
 \left| v_\ell \right|~=~\frac{m}{\sqrt{2 \left(a_0+a_\ell\right)}}\;,\qquad\text{for}~\;\ell=1,..,4\;,
\end{equation}
in agreement with the result of the classical minimization (cf.\ \Appref{app:potential}).

The analogous computation for the representation $\rep{72}_2$ yields a result which differs from \eqref{eq:Block_representants72} only by a global phase and 
thus gives no new VEVs if we take into account the freedom of a global $\U{1}_\text{Y}$ re--phasing.

Furthermore, solving \eqref{eq:consistency} for the permutation representation $\rep{27}$ results in 
\begin{equation}
\Phi_{\rep{27}}^\mathsf{T}~=~ \begin{pmatrix} \,\boxed{ G\,\phi_{27} }  ~~\end{pmatrix}\;,
\end{equation}
which only has a single block of which a representative is given by
\begin{equation}\label{eq:Block_representants27}
 \phi_{27}~=~ \begin{pmatrix} 0 \\ -\I \\ +\I \end{pmatrix}v_{27}\;.
\end{equation}
Plugging $\phi_{27}$ into the gradient of the potential we find that it is a stationary point if
\begin{equation}\label{eq:VEVstrength5}
  \left| v_{27} \right|~=~\frac{m\sqrt{3}}{\sqrt{4\,a_0+a_1+a_2+a_3+a_4}}\;,
\end{equation}
in agreement with the result obtained in the conventional way (cf.\ \Appref{app:potential}).

So far we have only investigated the largest subgroups of $E$ that allow for a non--trivial solution of \eqref{eq:consistency}.
In general, however, also subgroups of these subgroups allow for non--trivial solutions of \eqref{eq:consistency}, which can be less constraining, i.e.\ allow for more free parameters in $\Phi$.
We find that this is only the case if the corresponding subgroup is in the intersection of two or more larger subgroups of $E$ that allow for non--trivial solutions. 
This effect is, of course, to be expected because the corresponding solution $\Phi_\text{sub}$ has to accommodate all otherwise mutually exclusive solutions $\Phi_\text{parent}$ by fixing the additional parameters. 
If, instead, the subgroup is only contained in one larger group with non--trivial solution, we find that there are no additional parameters, i.e.\ the solution for the subgroup 
is identical to the solution of the parent group.

In our case, the potential only allows for precisely those VEVs which conserve the maximal possible subgroups of $E$ 
that allow for a non--trivial solution of \Eqref{eq:consistency}. 
All possible solutions with additional parameters retreat to the VEVs obtained from $\Phi_{\rep{27}}$ or $\Phi_{\rep{72}}$ when being plugged into the potential.
Due to the fact that the number of VEVs is bounded above, we can be sure to have found all VEVs of the potential once we have scanned over all non--trivial subgroups of \Equi.

Note, that our method does not provide us with an explanation of \textit{why} our potential realizes precisely those VEVs which preserve the largest subgroups of $E$.
In fact, our method does not provide any more information on the possible VEVs than what could be obtained by decomposing the representation of $H$ with respect to a particular subgroup and looking for trivial singlets.
What is new, however, is the insight that the corresponding subgroup is not only a subgroup of the symmetry group of the potential but, in fact, a subgroup of the complete group of all equivalence transformations. This provides us with more details regarding the structure of the VEVs.

Indeed, there is an interesting observation regarding the structure of the orbits of the stationary points. The \mbox{$\rep{72}_1$--plet} $\Phi_{\rep{72}}$ (and equivalently $\rep{72}_2$) decomposes under $G$ as
$\rep{72}_1=\rep{18}_1\oplus\rep{18}_2\oplus\rep{18}_3\oplus\rep{18}_4$, where $\rep{18}_\ell$ are permutation representations of $G$ corresponding to the individual blocks in $\Phi_{\rep{72}}$. 
The set of $\rep{18}$--plets does transform as a $\rep{4}$--plet under the action of the outer automorphism of $G$, $\Out(G)=\Sfour$.
Note, that this is the same transformation behavior as that of the four couplings $a_\ell$. 
As such, we observe that the stationary points $\Phi_{\rep{72}}$ do transform in the same representation as the couplings under the group of outer automorphisms,
while the stationary points $\Phi_{\rep{27}}$ transform as a trivial singlet.
The fact that VEVs under all allowed outer automorphisms either are invariant or transform in the same representation as the couplings themselves,
holds true for all cases that we have investigated (see below). 
Based on this observation we conjecture that this might be true in general.
In case this were true in general, this would constitute a remarkably easy method to calculate stationary points of potentials with outer automorphisms.

Let us comment on the applicability of our method to obtain the VEVs of other potentials.
It is clear that VEVs of any potential with a discrete symmetry or a discrete outer automorphism are solutions to consistency equations analogous to \eqref{eq:consistency}.
The explicit equations can, as in our case, be derived directly from the underlying symmetry group and the available outer automorphism transformations, 
where the latter depends on the representation content of a model as described in \Secref{sec:redundant}.
We have checked and confirmed that our method also enables us to find the VEVs of other potentials such as the pure \Dts potential without any other symmetries, 
which contains an additional cubic contraction of the triplets. 
Applying our method to the 3HDM potential with \A4 symmetry \cite{Degee:2012sk,Ivanov:2014doa} which allows for a \Z2 outer automorphism, only rough bounds on the form of the VEVs can be obtained.
If the outer automorphism group is trivial, as for instance in the 3HDM with $\Sfour$ symmetry, our method does not provide us with new constraints on the VEVs.

Since one of our motivations for this work was to study the origin of geometrical CP violation let us also comment on this.
In the case of our example, the calculability of the direction and phases of the VEVs is clearly attributed to the fact that the potential chooses the highest symmetric point not only with
respect to the symmetry group but also with respect to the outer automorphism group.
Even though we are not able to give sufficient conditions for the appearance of geometrical CP violation in general,
in our perception two conditions are necessary for the appearance of calculable phases.
Firstly, it seems necessary that the VEVs depend only on a small number of parameters, which is equivalent to $M$ in equation \eqref{eq:Mconsitency} having close to maximal rank.
In this way it is guaranteed that we can bring any VEV to the form $(v,0,0)$ by a (Higgs--)basis rotation that is independent of the couplings.
Secondly, in this new basis, there needs to be a CP transformation with fixed complex phases which is broken by this VEV.
Both of these conditions favor a large outer automorphism group.
Furthermore, the appearance of complex entries in the representation matrices of outer automorphisms is deeply related to the complexity of the Clebsch--Gordan coefficients of a group such that this consideration favors type I groups, according to the classification of 
\cite{Chen:2014tpa}.

\section{Summary and Conclusions}
\label{sec:Summary}

We have shown that outer automorphisms of the symmetry group of a model are relevant also beyond the usually considered C and P transformations. 
For a given model, only those outer automorphisms are possible which leave invariant the set of all present representations. 
We have termed those transformations equivalence transformations.
All other outer automorphisms are broken explicitly and maximally.
As subsets, equivalence transformations contain C, P, or CP transformations and, as the trivial case, also symmetry transformations.

We have shown that the effect of non--trivial equivalence transformations on a theory is to map couplings to different values, i.e.\ the theory to a different point in the parameter space.
This may, for example, manifest itself as a permutation of couplings.
Moreover, it implies that the according field redefinitions have the same effect as a shift of the couplings.
Because field redefinitions cannot matter physically we see that the respective transformations of the couplings cannot matter physically. 
For the complete discussion of the physical phenomenology of a model
it is, thus, sufficient to consider only a restricted region of the parameter space which is
related to the complete parameter space by equivalence transformations.
Stated in other terms, the possibility of equivalence transformations signals physical degeneracies in the parameter space of a theory.

Moreover, we have shown that stationary points always transform in certain representations of the group of equivalence transformations.
Thus, given a stationary point one may employ outer automorphism transformations to obtain others.
Exploiting this further, we were able to derive a set of homogeneous linear equations which constrain the phases and directions of stationary points.
Curiously, for all examples that we have investigated, we find that stationary points either are invariant or transform in the same representation as the couplings themselves.
We conjecture that this might be true in general. Proven true, this could explain why minima of potentials often are located at symmetry enhanced points.
In this respect, we think that a deeper investigation of the subject from a more mathematical point of view would certainly be worthwhile.

For the three Higgs doublet model with \Dts symmetry we have derived the complete group of equivalence transformations and explicitly confirmed that for a suitable 
choice of parameters CP is spontaneously violated by the VEVs of the Higgs fields.
The parameter independent directions and relative phases of the VEVs can be understood to originate from two facts. 
Firstly, the homogeneous linear equations derived from the symmetry group and representation content of the potential 
are so restrictive on the most symmetric VEV candidates as to completely fix their direction and relative phases.
Secondly, the potential realizes precisely those, most symmetric VEVs.
We have found that all stationary points which can be the global minimum of the potential are part of a quadruplet under the outer automorphism transformations.
As such, all possible VEVs conserve isomorphic subgroups of the symmetry group and of the group of outer automorphisms.
To be clear, this implies that VEVs of the form $v\,(1,0,0)$ are not distinguished in their phenomenology, e.g.\ concerning the geometrical violation of CP, from VEVs of the form $v\,(\omega,1,1)$.
Their physical implications only become different if the equivalence transformations which relate them are prohibited explicitly.
This can only happen if a sector with fields in additional representations, for example \Dff doublets, is included in the model.

Even though we are not able to formulate generally valid necessary and sufficient conditions, we argue that the appearance of geometrical CP is favored by groups with complex Clebsch--Gordan coefficients and a large outer automorphism group.

Furthermore, we think that it would be interesting to explore the implications of equivalence transformations on the shape of the so--called ``orbit space'' \cite{Ivanov:2010ww,Degee:2012sk} (not to be confused with the orbits used in \Secref{sec:calculableVEVs}) which has been used to minimize potentials in a geometrical way.

Finally, let us remark that our discussion is, in principle, not limited to discrete groups.
As such it would certainly be worthwhile to investigate what other theories allow for non--trivial equivalence transformations.

\section*{Acknowledgments}
It is a pleasure to thank Michael Ratz for enlightening discussions and useful comments on the manuscript.
Furthermore, we would like to thank Mu--Chun Chen and Thomas Rauh for useful discussions, and Igor Ivanov and Celso Nishi for sharing an early version of their manuscript \cite{Ivanov:2014doa}. We thank Igor Ivanov and Ermal Rrapaj for useful comments.
This research was supported in parts by the DFG cluster of excellence ``Origin and Structure of the Universe'', the DFG Research Grant ``Flavor and CP in supersymmetric extensions of the Standard Model'',
the DFG Graduiertenkolleg 1054 ``Particle Physics at the Energy Frontier of New Phenomena'' and the TUM Graduate School.
This research was done in the context of the ERC Advanced Grant project ``FLAVOUR'' (267104).

\appendix
\section{Group theory of \texorpdfstring{$\boldsymbol{\Dts}$}{Delta(27)} and \texorpdfstring{$\boldsymbol{\Dff}$}{Delta(54)}}
\label{app:Delta27}

In this Appendix we gather the group theory relevant to the present work.
Details on the group \Dts can be found in \cite{Chen:2014tpa} from which we also adopt the notation and basis convention.
\Dts and \Dff are included in the catalogue of
GAP \cite{GAP4} as $\mathrm{SG}(27,3)$ and $\mathrm{SG}(54,8)$, respectively.

A possible presentation for the group \Dff is given by the operations $\mathsf{A}$, $\mathsf{B}$, and $\mathsf{C}$, where
\begin{equation}\label{eq:presentationG}
 \mathsf{A}^3~=~\mathsf{B}^3~=~\mathsf{C}^2~=~\left(\mathsf{A}\,\mathsf{B}\right)^3~=~\left(\mathsf{A}\,\mathsf{C}\right)^2~=~\left(\mathsf{B}\,\mathsf{C}\right)^2~=~\mathsf{e}\;.
\end{equation}
The conjugacy classes are given by
\begin{align}
C_{1a}& : \mathsf{\{e \}\;,}               &  \nonumber\\
C_{3a}& : \mathsf{\{A, A^2, BAB^2, B^2AB, BA^2B^2, B^2A^2B \}\;,} &\nonumber\\
C_{3b}& : \mathsf{\{B, B^2, ABA^2, A^2BA, AB^2A^2, A^2B^2A \}\;,} &\nonumber\\
C_{3c}& : \mathsf{\{AB^2, A^2B, BA^2, B^2A, ABA, BAB \}\;, }      &\nonumber\\
C_{3d}& : \mathsf{\{AB, BA, A^2B^2, B^2A^2, AB^2A, A^2BA^2 \}\;,} &\nonumber\\
C_{2a}& : \mathsf{\{C, AC, A^2C, BC, B^2C, ABAC, BABC, A^2BA^2C, AB^2AC \}\;,} &\nonumber\\
C_{6a}& : \mathsf{\{BAC, A^2BC, AB^2C, B^2A^2C, B^2ABC, BA^2B^2C, ABA^2C, A^2B^2AC, AB^2ABAC \}\;,} &\nonumber\\
C_{6b}& : \mathsf{\{ABC, BA^2C, B^2AC, A^2B^2C, A^2BAC, BAB^2C, AB^2A^2C, B^2A^2BC, BA^2BABC \}\;,} &\nonumber\\
C_{3e}& : \mathsf{\{AB^2ABA \}\;,}\qquad C_{3f}: \mathsf{\{BA^2BAB\}\;.}&
\label{eq:ccs}
\end{align}
The non--trivial irreducible representations consist of the real representations $\rep[_1]{1}$ and $\rep[_i]{2}$ (with $1\le i \le 4$)
and the complex representations $\rep[_1]{3}$ and $\rep[_2]{3}$ and their respective conjugates.  The character table is
given in \ref{tab:Delta54char}.
\renewcommand{\arraystretch}{1.}
\begin{table}[t!]
\centering
\resizebox{\textwidth}{!}{\begin{tabular}{c|rrrrrrrrrr|}
                      &  $C_{1a}$ & $C_{3a}$ & $C_{3b}$ & $C_{3c}$ & $C_{3d}$ & $C_{2a}$ & $C_{6a}$ & $C_{6b}$ & $C_{3e}$ & $C_{3f}$  \\
                      &  1 &  6 &  6 &  6 &  6 &  9 &  9 &  9 &   1 &  1  \\
\Dff         & $\mathsf{e}$ & $\mathsf{A}$  & $\mathsf{B}$  & $\mathsf{ABA}$ & $\mathsf{AB}$ & $\mathsf{C}$  & $\mathsf{ABC}$       & $\mathsf{BAC}$   & $\mathsf{AB^2ABA}$ & $\mathsf{BA^2BAB}$ \\
\hline
 $\rep[_0]{1}$       & $1$ & $1$  & $1$  & $1$  & $1$  & $1$  & $1$         & $1$         & $1$         & $1$        \\
 $\rep[_1]{1}$       & $1$ & $1$  & $1$  & $1$  & $1$  & $-1$ & $-1$        & $-1$        & $1$         & $1$        \\
 $\rep[_1]{2}$       & $2$ & $2$  & $-1$ & $-1$ & $-1$ & $0$  & $0$         & $0$         & $2$         & $2$        \\
 $\rep[_2]{2}$       & $2$ & $-1$ & $2$  & $-1$ & $-1$ & $0$  & $0$         & $0$         & $2$         & $2$        \\
 $\rep[_3]{2}$       & $2$ & $-1$ & $-1$ & $2$  & $-1$ & $0$  & $0$         & $0$         & $2$         & $2$        \\
 $\rep[_4]{2}$       & $2$ & $-1$ & $-1$ & $-1$ & $2$  & $0$  & $0$         & $0$         & $2$         & $2$        \\
 $\rep[_1]{3}$       & $3$ & $0$  & $0$  & $0$  & $0$  & $1$  & $\omega^2\hspace{-5pt}$  & $\omega$    & $3\omega$ & $3\omega^2\hspace{-5pt}$  \\
 $\rep[_1]{\bar{3}}$ & $3$ & $0$  & $0$  & $0$  & $0$  & $1$  & $\omega$    & $\omega^2\hspace{-5pt}$  & $3\omega^2\hspace{-5pt}$ & $3\omega$  \\
 $\rep[_2]{3}$       & $3$ & $0$  & $0$  & $0$  & $0$  & $-1$ & $-\omega^2\hspace{-5pt}$ & $-\omega$   & $3\omega$ & $3\omega^2\hspace{-5pt}$  \\
 $\rep[_2]{\bar{3}}$ & $3$ & $0$  & $0$  & $0$  & $0$  & $-1$ & $-\omega$   & $-\omega^2\hspace{-5pt}$ & $3\omega^2\hspace{-5pt}$ & $3\omega$  \\
\hline
\end{tabular}}
\caption{Character table of \Dff. We define $\omega:=\mathrm{e}^{2\pi\,\I/3}$. 
The conjugacy classes (c.c.) are labeled by the order of their elements and a
letter.  The second line gives the cardinality of the corresponding c.c.\ and
the third line gives a representative of the c.c.\ in the presentation specified
in the text.}
\label{tab:Delta54char}
\end{table}

For the triplets $\rep[_{1,2}]{3}$ we use the representation matrices
\begin{equation}\label{eq:delta54gens}
 A~=~\begin{pmatrix} 0 & 1 & 0 \\ 0 & 0 & 1 \\ 1 & 0 & 0 \end{pmatrix}\;, 
 \qquad B\,=\,\begin{pmatrix} 1 & 0 & 0 \\ 0 & \omega & 0 \\ 0 & 0 & \omega^2 \end{pmatrix}\;,
 \qquad C\,=\,\pm\begin{pmatrix} 1 & 0 & 0 \\ 0 & 0 & 1 \\ 0 & 1 & 0 \end{pmatrix}\;,
\end{equation}
and for $\rep[_{1,2}]{\bar{3}}$ the respective complex conjugate matrices. We have checked explicitly that this is the correct form for $C$ contrary to earlier statements in the literature.

The group \Dts is a normal subgroup of \Dff which can be obtained by dropping the generator $\mathsf{C}$.
The restriction of the conjugation map $\operatorname{conj}(\mathsf{C})$ from \Dff to \Dts leads to an outer automorphism of \Dts, 
which acts as an exchange of all singlet representations with their respective complex conjugates.%
\footnote{Since the Higgs potential \eqref{eq:Invariants} is built only out of contractions which are symmetric under this exchange, the symmetry of the Higgs potential is not \Dts but \Dff.}
The real \Dff--doublets thus can be obtained from the pairs of mutually complex conjugate one--dimensional representations of \Dts as 
$\rep[_1]{2}=(\rep[_1]{1},\rep[_2]{1})$, $\rep[_2]{2}=(\rep[_3]{1},\rep[_6]{1})$, $\rep[_3]{2}=(\rep[_4]{1},\rep[_8]{1})$, and $\rep[_4]{2}=(\rep[_5]{1},\rep[_7]{1})$.
The triplet representation is sent to itself under the automorphism with a matrix representation of $\mathsf{C}$ acting on the 
triplet as $C$, given in \eqref{eq:delta54gens}. 
Therefore, all outer automorphisms of \Dts are also available at the level of \Dff, where the ones related to $\mathsf{C}$, however, are inner automorphisms and, therefore, automatically realized.

The Clebsch--Gordan coefficients of \Dff relevant to this work are given by
\begin{subequations}\label{eq:Delta54CGs}
\begin{align}
  \left(x_{\rep[_i]{2}}\otimes y_{\rep[_i]{2}}\right)_{\rep[_0]{1}} & ~=~ \frac{1}{\sqrt{2}}\left(x_1\, y_2 + x_2\, y_1 \right)\;,\nonumber\\
	\left(x_{\rep[_i]{3}}\otimes y_{\rep[_i]{\bar{3}}}\right)_{\rep[_0]{1}} & ~=~ \frac{1}{\sqrt{3}}\left(x_1\, \bar{y}_1 + x_2\, \bar{y}_2 + x_3\, \bar{y}_3 \right)\;,\nonumber\\
  \left(x_{\rep[_i]{3}}\otimes y_{\rep[_i]{\bar{3}}}\right)_{\rep[_1]{2}} & ~=~ 
  \frac{1}{\sqrt{3}}\begin{pmatrix}x_1\, \bar{y}_2 + x_3\, \bar{y}_1 + x_2\, \bar{y}_3 \\ x_2\, \bar{y}_1 + x_1\, \bar{y}_3 + x_3\, \bar{y}_2 \end{pmatrix}\;,\nonumber\\
  \left(x_{\rep[_i]{3}}\otimes y_{\rep[_i]{\bar{3}}}\right)_{\rep[_2]{2}} & ~=~ 
  \frac{1}{\sqrt{3}}\begin{pmatrix}x_1\, \bar{y}_1 + \omega\, x_2\, \bar{y}_2 + \omega^2\, x_3\, \bar{y}_3 \\ x_1\, \bar{y}_1 + \omega^2\, x_2\, \bar{y}_2 + \omega\, x_3\, \bar{y}_3 \end{pmatrix}\;,\nonumber\\
  \left(x_{\rep[_i]{3}}\otimes y_{\rep[_i]{\bar{3}}}\right)_{\rep[_3]{2}} & ~=~ 
  \frac{1}{\sqrt{3}}\begin{pmatrix}x_2\, \bar{y}_3 + \omega\, x_3\, \bar{y}_1 + \omega^2\, x_1\, \bar{y}_2 \\ \omega\, x_2\, \bar{y}_1 + x_3\, \bar{y}_2 + \omega^2\, x_1\, \bar{y}_3 \end{pmatrix}\;,\nonumber\\
  \left(x_{\rep[_i]{3}}\otimes y_{\rep[_i]{\bar{3}}}\right)_{\rep[_4]{2}} & ~=~ 
  \frac{1}{\sqrt{3}}\begin{pmatrix}\omega^2\, x_2\, \bar{y}_1 + x_3\, \bar{y}_2 + \omega\, x_1\, \bar{y}_3 \\ x_2\, \bar{y}_3 + \omega^2\, x_3\, \bar{y}_1 + \omega\, x_1\, \bar{y}_2 \end{pmatrix}\;.
\end{align}
\end{subequations}
CGs for other contractions can be found in \cite{Escobar:2011mq}, but one should be aware of the fact that we use a different labeling for the representations.
As one can check by computing the twisted Frobenius--Schur indicators for all automorphisms, \Dff is of type I according to the classification of \cite{Chen:2014tpa}.

\section{Minimization of the potential}\label{app:potential}

We want to give details of the traditional minimization procedure of the Higgs potential in this appendix.
In order to have a potential which is bounded below the parameters have to satisfy the conditions  
\begin{equation}\label{eq:physicality}
 0~<~\lambda_1 \qquad\text{and}\qquad~0~<~\lambda_1+\lambda_{23}+2\,\lambda_4\,\cos\left[2\pi/3 + (\Omega\mod 2\pi/3)\right]\;,
\end{equation}
where $\lambda_{23}:=\lambda_2+\lambda_3$.
The VEVs are solutions to
\begin{equation}
 0~\stackrel{!}{=}~\left.\frac{\partial\,V}{\partial |H_i|}\right|_{H_i=\langle H_i \rangle}~~~{\rm and}~~~0~\stackrel{!}{=}~\left.\frac{\partial\,V}{\partial\varphi_i}\right|_{H_i=\langle H_i \rangle}\;,
 \label{eq:stationary}
\end{equation}
where we assume that parameters are such that the (electric) charge is conserved and parametrize the VEVs as in \eqref{eq:VEVparametrization}. This will be justified 
a posteriori, cf.\ \Appref{app:charge_conservation}. Among the solutions to \eqref{eq:stationary} there are all types of stationary points 
and the true global minima have been identified by explicitly computing the value of the potential at the stationary points, as outline below.

Let us first focus on the magnitude of the stationary points. Defining 
\begin{equation}
\bar{\theta}_i~:=~-2\,\varphi_i+\varphi_j+\varphi_k+\Omega\;,~~\text{for}~i \neq j\neq k\neq i=1,2,3\;,
\end{equation}
the first condition of \eqref{eq:stationary} leads to
\begin{equation}\label{eq:magnitude_condition}
 0~\stackrel{!}{=}~-m^2\,v_1+2\,\lambda_1\,{v_1}^3 + \lambda_{23}\,v_1\left({v_2}^2+{v_3}^2\right) + \lambda_4\,v_2\,v_3\left[2\,v_1\,\cos\bar{\theta}_1 + v_2\,\cos\bar{\theta}_2 + v_3\,\cos\bar{\theta}_3\right]\;,
\end{equation}
and two more equations obtained by cyclic permutation of the indices of $v_i$ and $\bar{\theta}_i$. 
In order to determine $v_i$ we have to solve this system of three coupled cubic equations. There are at most 27 real solutions for $|\langle H \rangle|=(v_1,v_2,v_3)$. 
Because the equations have the permutation symmetry stated above, also the possible solutions will obey this symmetry and we only have to investigate a substantially smaller set of solutions. 
Also, one should keep in mind that we are interested only in real and positive solutions.

The solutions split in four categories. The trivial solution $|\langle H \rangle|=h^{(0)}:=(0,0,0)$ is always a local maximum of the potential. 
Furthermore, there are $6 (= 3$ permutations $\times$ 2 possible signs$)$ solutions of the type $h^{(1)}:=(v^{(1)},0,0)$, $3\times2^2=12$ solutions of the type $h^{(2)}:=(v^{(2)},v^{(2)},0)$ 
and $1\times2^3$ solutions of the type $h^{(3)}:=(v^{(3)},v^{(3)},v^{(3)})$. All these possibilities, if simply imposed as an Ansatz, can be shown to be solutions to the extremization condition.
Since all the possibilities sum up to 27, we can be sure that no solutions have been missed.

The respective magnitudes of the solutions are given by
\begin{equation}\label{eq:stationary_points}
\begin{split}
 v^{(1)}~=~&\sqrt{\frac{m^2}{2\,\lambda_1}}\;, \quad v^{(2)}~=~\sqrt{\frac{m^2}{2\,\lambda_1+\lambda_{23}}}\;, \quad \text{and} \\
 v^{(3)}~=~&\sqrt{\frac{m^2}{2}}\left[ \lambda_1+\lambda_{23}+2\,\lambda_4\,\cos\bar{\theta}_1\right]^{-1/2}\;.
 \end{split}
\end{equation}
The dependence on one particular $\bar{\theta}_i$ in the last relation should not lead to confusion. We will show below that in case of $h^{(3)}$ the phase dependent potential warrants
that $\bar{\theta}_1=\bar{\theta}_2=\bar{\theta}_3=\left[2\pi/3 + (\Omega\mod 2\pi/3)\right]$ at the stationary point.

In the case of $h^{(2)}$, one of the equations \eqref{eq:magnitude_condition} gives a condition on the possible phases of the VEV, constraining the relative phase between the two entries to be $\pm\pi/3$ or $\pi$.

Let us now investigate the stationary points of the phase dependent potential $V_I$. The second condition of \eqref{eq:stationary} leads to
\begin{equation}
0~\stackrel{!}{=}~\lambda_4\,v_1\,v_2\,v_3\left[ 2\,v_1\,\sin{\bar{\theta}_1}-v_2\,\sin{\bar{\theta}_2}-v_3\,\sin{\bar{\theta}_3} \right]\;,
\label{eq:phase_condition}
\end{equation}
and two more equations which are again obtained by cyclic permutation of the indices of $v_i$ and $\bar{\theta}_i$. Obviously, 
for the stationary points of the form $h^{(1)}$ and $h^{(2)}$ all three equations are trivial and hence there is no constraint on the possible phases from \eqref{eq:phase_condition}.

In contrast to that, we will see that in the case of $h^{(3)}$ possible (relative) phases of the VEVs will be fixed to discrete values. In order to investigate this case, 
we take $v_1=v_2=v_3$ and therefore obtain from \eqref{eq:magnitude_condition} and \eqref{eq:phase_condition} the relations
\begin{equation}
 \cos{\bar{\theta}_1}~=~\cos{\bar{\theta}_2}~=~\cos{\bar{\theta}_3}\qquad \text{and} \qquad \sin{\bar{\theta}_1}~=~\sin{\bar{\theta}_2}~=~\sin{\bar{\theta}_3}\;,
\end{equation}
which imply that $\bar{\theta}_1=\bar{\theta}_2=\bar{\theta}_3$. To obtain the respective values for the $\varphi_i$ it is convenient to fix $\varphi_3=0$.
This is always possible since the phases $\varphi_i$ are meaningful only relative to each other because an overall global phase of $\langle H \rangle$ can always be removed by a global hypercharge rotation.
Doing this, we obtain the relation 
\begin{equation}
 \varphi_1~=~\varphi_2~=~0\mod 2\pi/3\;.
\end{equation}
This implies that here are nine possible combinations of discrete phases each corresponding to one stationary point.
In \Tabref{tab:stationary_phases} we list all the possibilities for $\varphi_i$, the corresponding value of $\bar{\theta}_i$ and the value of the potential at the stationary point.
Which stationary point is a global minimum  critically depends on the value of $\Omega$ as also depicted in \Figref{fig:potential}.

We find that the nine stationary points may be classified into three types I, II, and III.
Depending on the value of $\Omega$ the potential can have exactly six energy--degenerate global minima which are of 
type I and II ($\Omega=2\pi/3$), type I and III ($\Omega=4\pi/3$), or type II and III ($\Omega=0$). 
For all other values of $\Omega$, there are exactly three energy--degenerate minima of the same type, where the type is determined by the tertial in which $\Omega$ lies.
The stationary points of each type are physically equivalent because they are part of the same group orbit.

\renewcommand{\arraystretch}{1.3}
\begin{table}[t]
\begin{center}
\begin{tabular}{lcccccc}
       & $(\varphi_1,\varphi_2)$ & $\#$ & $\bar{\theta}_{i=1,2,3}$ & $V_I(\Omega)/{v^{(3)}}^4$ & $\langle H \rangle/v^{(3)}$ & global min.\ for \\ 
\hline
\multirow{2}{*}{Type I} & $(0,0)$ & 1 & \multirow{2}{*}{$\Omega$}  & \multirow{2}{*}{$\propto\cos{\Omega}$} & \multirow{2}{*}{$(1,1,1)^\mathsf{T}$} & \multirow{2}{*}{$\frac{2\pi}{3}\leq\Omega\leq\frac{4\pi}{3}$} \\
       & $\left(\frac{2\pi}{3},\frac{4\pi}{3}\right)$ & 2 &           &    &     &     \\
\multirow{2}{*}{Type II} & $\left(0,\frac{2\pi}{3}\right)$ & 2 & \multirow{2}{*}{$\frac{2\pi}{3}+\Omega$}  & \multirow{2}{*}{$\propto\cos\left(\frac{2\pi}{3}+\Omega\right)$} 
       & \multirow{2}{*}{$(\omega,1,1)^\mathsf{T}$} & \multirow{2}{*}{$0\leq\Omega\leq\frac{2\pi}{3}$}\\
       & $\left(\frac{4\pi}{3},\frac{4\pi}{3}\right)$ & 1 &           &      &    &    \\
\multirow{2}{*}{Type III} & $\left(0,\frac{4\pi}{3}\right)$ & 2 & \multirow{2}{*}{$\frac{4\pi}{3}+\Omega$}  & \multirow{2}{*}{$\propto\cos\left(\frac{4\pi}{3}+\Omega\right)$} 
       & \multirow{2}{*}{$(\omega^2,1,1)^\mathsf{T}$} & \multirow{2}{*}{$\frac{4\pi}{3}\leq\Omega\leq2\pi$}\\
       & $\left(\frac{2\pi}{3},\frac{2\pi}{3}\right)$ & 1 &           &       &    &   \\       
\end{tabular}
\caption{List of stationary points of the phase dependent potential $V_I(\varphi_i, \Omega)$ in dependence of $\varphi_1$ and $\varphi_2$ relative to $\varphi_3$. 
The third column gives the multiplicity of the stationary point (for permutations of the value of $\varphi_1$ and $\varphi_2$). In the fourth, fifth, and sixth column we list the value of 
$\bar{\theta}_1=\bar{\theta}_2=\bar{\theta}_3$ at the stationary point, the corresponding value of the potential, as well as an example for the phases of the complete Higgs triplet VEV, respectively. The last column gives the region in 
$\Omega$--parameter space in which the stationary points of the respective type are the global minima of the potential (cf.\ Fig.\ref{fig:potential}).}
\label{tab:stationary_phases}
\end{center}
\end{table}

From the preceding discussion it follows that at the minimum of the potential we can always write 
\begin{equation}
\bar{\theta}_i~=~2\pi/3+(\Omega\mod2\pi/3)\;.
\end{equation}
Using this in \Eqref{eq:stationary_points} and comparing the value of the overall potential at the stationary points we find that in case 
\begin{equation}
 2\,\lambda_1 ~>~\lambda_{23} + 2\,\lambda_4\,\cos\left[2\pi/3 + (\Omega\mod 2\pi/3)\right]\;,
 \label{eq:minimum_condition}
\end{equation}
the global minima of the potential is given by VEVs of the form $h^{(3)}$ with possible phases determined by the value of $\Omega$. 
If \eqref{eq:minimum_condition} is violated, the global minima will be of the form $h^{(1)}$. 
The stationary points $h^{(2)}$ will never be global minima of the potential, even though in the case $\lambda_4\neq0$ they can be local ones.

A comment is in order regarding the two different parametrizations used in this work. Using \eqref{eq:parametrization} one can show with some trigonometry that the conditions \eqref{eq:physicality} 
coincide with 
\begin{equation}\label{eq:physicalityA}
 0~<~a_0+a_\ell\;, \qquad\text{for}~\;\ell=1,..,4\;,
\end{equation}
which one would obtain from \eqref{eq:VEVstrength}. The same is true for the magnitude of the stationary points, \Eqref{eq:stationary_points} and \eqref{eq:VEVstrength} or \eqref{eq:VEVstrength5}, respectively.

\subsection{Charge--breaking and charge--conserving vacua} \label{app:charge_conservation}

Unlike in the case of models with one or two Higgs doublets, where an existing charge--conserving vacuum is automatically also the global minimum of the potential \cite{Barroso:2005da,DiazCruz:1992uw},
in a three Higgs doublet model it has to be checked explicitly whether the global minimum of the potential is really charge--conserving \cite{Barroso:2006pa}.

In case of a type $\rm{IV}$ global minimum, only one of the doublets acquires a VEV and, hence, it is always possible to show that there cannot be any lower-lying charge breaking global minimum \cite{Barroso:2006pa}.
The fact that VEVs of type $\rm{I-III}$ are, at the level of the potential, equivalent to the VEV of type $\rm{IV}$ implies that any of the VEVs can be brought to the form $\rm{IV}$ by an appropriate basis rotation.
Again following the argumentation of \cite{Barroso:2006pa}, this shows that also global minima of the type $\rm{I-III}$ are charge conserving.

\section{Proof of Eq.\ \texorpdfstring{(\ref{eq:additionalVEV})}{Eq.} }
\label{app:VEVtrafo}

Let us assume that the potential $V(H,\alpha)$ has a stationary point $\phi(\alpha):=\langle H \rangle$, that is
\begin{equation}\label{eq:StationaryPoint}
  \left.\frac{\partial\,V(H,\alpha)}{\partial\,H_i}\right|_{H=\phi(\alpha)} ~=~ \left.\frac{\partial\,V(H,\alpha)}{\partial\,H_i^*}\right|_{H=\phi(\alpha)} ~=~ 0 \qquad \forall~i\;.
\end{equation}
Also, we assume that the potential allows for equivalence transformations, i.e.\ it fulfills the relation
\begin{equation}\label{eq:ETPotential}
 V(H',\alpha)~=~V(H,\alpha')\;,
\end{equation}
where we denote by $\alpha'$ the transformed parameters, and by $H'=UH$ or $H'=UH^*$, depending on the case at hand, the equivalence transformed fields.
In order to prove \eqref{eq:additionalVEV}, we have to show that
\begin{align}
  & \left.\frac{\partial\,V(H,\alpha)}{\partial\,H_i}\right|_{H=U\,\phi\left(\alpha'\right)}~=
  ~\left.\frac{\partial\,V(H,\alpha)}{\partial\,H_i^*}\right|_{H=U\,\phi\left(\alpha'\right)}~=~0 \qquad \forall~i\\
  \intertext{or}
  & \left.\frac{\partial\,V(H,\alpha)}{\partial\,H_i}\right|_{H=U\,\left(\phi\left(\alpha'\right)\right)^*}~=
  ~\left.\frac{\partial\,V(H,\alpha)}{\partial\,H_i^*}\right|_{H=U\,\left(\phi\left(\alpha'\right)\right)^*}~=~0 \qquad \forall~i\;,
\end{align}
respectively.

Let first $H'=UH$.
Then
\begin{equation}
\begin{split}
  0 ~\stackrel{\eqref{eq:StationaryPoint}}{=}&~\left.\frac{\partial\,V(H,\alpha')}{\partial\,H_i}\right|_{H=\phi\left(\alpha'\right)}~\stackrel{\eqref{eq:ETPotential}}{=}~
  \left.\frac{\partial\,V(UH,\alpha)}{\partial\,H_i}\right|_{H=\phi\left(\alpha'\right)}~=~ \\
  ~\stackrel{\phantom{\eqref{eq:StationaryPoint}}}{=}&~\left.\frac{\partial\,V(H,\alpha)}{\partial\,H_j}\right|_{H=U\,\phi\left(\alpha'\right)} \cdot U_{ji}\;,
\end{split}
\end{equation}
where the last equality follows from the chain rule.
Together with the analogous equation for the derivative with respect to the conjugate field, and by noting that $U$ is invertible, this proves the assertion for the first case.

The second case is only slightly more involved because one has to take care of the complex conjugation.
Let now $H'=UH^*$.
Then
\begin{equation}
\begin{split}
  0 ~\stackrel{\eqref{eq:StationaryPoint}}{=}&~ \left.\frac{\partial\,V(H,\alpha')}{\partial\,H_i}\right|_{H=\phi\left(\alpha'\right)}~\stackrel{\eqref{eq:ETPotential}}{=}~
  \left.\frac{\partial\,V(UH^*,\alpha)}{\partial\,H_i}\right|_{H=\phi\left(\alpha'\right)}~=~ \\
  ~\stackrel{\phantom{\eqref{eq:StationaryPoint}}}{=}&~\left.\frac{\partial\,V(H,\alpha)}{\partial\,H_j}\right|_{H=U\,\left(\phi\left(\alpha'\right)\right)^*} \cdot 0 ~+~  \left.\frac{\partial\,V(H,\alpha)}{\partial\,H_j^*}\right|_{H=U\,\left(\phi\left(\alpha'\right)\right)^*} \cdot U_{ji}^*\;,
\end{split}
\end{equation}
where the last equality follows from the chain rule (for complex derivatives). 
Together with the analogous equation for the derivative with respect to the conjugate field, and by noting that $U^*$ is invertible, this proves the assertion for the second case.

\section{Permutation representations}\label{app:permutations}\label{App:VEVCalculation}

Let $\Equi$ be a group with subgroup $\Equi_\phi$. Let the group act via left--multiplication on the coset $E/E_\phi$.
This defines a permutation representation. The explicit permutation matrix $(\Pi_{\mathsf{P}})^{-1}$ of a group element $\mathsf{P}\in E$ in this representation
can be computed with the following GAP \cite{GAP4} code:

\small
\begin{verbatim}
action:=ActionHomomorphism(E,RightCosets(E,E_phi),OnRight);;
Pi_P_inverse:=Image(action,P);
\end{verbatim}
\normalsize
Using this, the minimal generating set of the permutation representation $\rep{72}_1$ (in cycles) is given by
\begin{equation}
\begin{split}
 (\Pi_{\mathsf{P}}^{\rep{72}_1})^{-1}~:=&~(2,9,5)(4,13,33)(6,17,14)(7,19,15)(8,22,47)(10,26,24)(11,28,53) \\ 
 &~ (12,31,55)(16,38,58)(18,40,59)(20,42,60)(21,44,62)(23,46,65) \\ 
 &~ (25,49,66)(27,51,67)(29,45,64)(30,54,69)(32,57,35)(37,52,68) \\
 &~ (43,61,71)(48,56,70)(50,63,72)\;, \\
 (\Pi_{\mathsf{Q}}^{\rep{72}_1})^{-1}~:=&~(1,59, 8,56,26,72,37,44)( 2,33,18,63,36,68,30,46) \\ 
 &~ ( 3,60,25,38, 9,64,35,61)( 4,42,19,70,11,40,17,58) \\
 &~ ( 5,62,12,28,41,67,48,22)(6,65,29,52, 7,69,50,32) \\
 &~ (10,53,20,45,34,57,23,54)(13,39,55,16,49,24,71,27) \\
 &~ (14,47,21,51,15,66,43,31)\;.
\end{split}
\end{equation}
The minimal generating set of the permutation representation $\rep{72}_2$ is given by
\begin{equation}
\begin{split}
 (\Pi_{\mathsf{P}}^{\rep{72}_2})^{-1}~:=&~(2,9,5)(4,13,33)( 6,17,14)( 7,19,15)( 8,22,47)(10,26,24)(11,28,54)\\ 
 &~(12,31,55)(16,38,58)(18,40,59)(20,42,60)(21,44,63)(23,48,66)\\ 
 &~(25,50,67)(27,52,68)(29,45,64)(30,46,65)(32,57,35)(37,53,69)\\ 
 &~(43,61,71)(49,56,70)(51,62,72)\;, \\
 (\Pi_{\mathsf{Q}}^{\rep{72}_2})^{-1}~:=&~( 1,58,29,53,23,54,42,18)( 2,63,22,55,33,62,31,26)\\
 &~( 3,47,17,52, 4,71,37,49)( 5,60,21,68,24,38,27,61)\\
 &~( 6,56,13,57,15,64,48,41)( 7,59,30,36, 8,72,35,32)\\
 &~( 9,50,12,70,34,46,20,51)(10,40,25,44,14,67,45,43)\\
 &~(11,66,19,65,28,69,16,39)\;.
\end{split}
\end{equation}
The minimal generating set of the permutation representation $\rep{27}$ is given by
\begin{equation}
\begin{split}
 (\Pi_{\mathsf{P}}^{\rep{27}})^{-1}~:=&~( 1, 4,12)( 2, 8,20)( 3, 9,21)( 5,15,24)( 6,16,11) \\
 &~ ( 7,17,19)(10,23,25)(13,27,14)(18,22,26)\;, \\
 (\Pi_{\mathsf{Q}}^{\rep{27}})^{-1}~:=&~( 1, 8,25,24,14,27,26,13)( 2,22,15, 9,11,19,16, 4) \\
 &~( 3,10,21,12,20,17, 5, 6)(18,23)\;.
\end{split}
\end{equation}

\bibliography{SymmetriesOfSymmetriesbib}
\bibliographystyle{NewArXiv}

\end{document}